\documentclass[11pt,draftclsnofoot,peerreview,onecolumn,notitlepage]{IEEEtran}
\IEEEoverridecommandlockouts

\usepackage{cite}
\usepackage{ieeefig}
\usepackage{amsfonts}
\usepackage{epsfig,color}
\usepackage{graphicx}
\usepackage[caption=false,font=footnotesize]{subfig}
\usepackage{stfloats}
\usepackage{amsmath}
\usepackage{paralist}
\usepackage{color}

\newtheorem{theorem}{\bf Theorem}

\newtheorem{remark}{\bf Remark}
\newtheorem{fact}{\bf Fact}
\newtheorem{definition}{\bf Definition}

\newcommand{\define}    {\stackrel{\triangle}{=}}  
\newcommand{\snr}{\mathsf{SNR}} \newcommand{\inr}{\mathsf{INR}}
\newcommand{\snrs}{\mathsf{s}{\sf n}{\sf r}}
\newcommand{\Cpp}{\mathbf{\mathsf{C}}}

\begin{document}
%
\title{\LARGE{Independent signaling achieves the capacity region of the Gaussian interference channel with common information to within one bit} }

\author{Chinmay S.~Vaze %
        and~Mahesh K.~Varanasi
\thanks{The authors are with the Department of Electrical, Computer, and Energy
Engineering, University of Colorado, Boulder, CO 80309-0425 USA
(e-mail: {Chinmay.Vaze, varanasi}@colorado.edu). The material in this paper was presented in part
at the 2011 International Symposium on Information Theory (ISIT), St. Petersburg, Russia.} }

%



\maketitle

\begin{abstract}
The interference channel with common information (IC-CI) consists of two transmit-receive pairs that communicate over a common noisy medium. Each transmitter has an individual message for its paired receiver, and additionally, both transmitters have a common message to deliver to both receivers. In this paper, through explicit inner and outer bounds on the capacity region, we establish the capacity region of the Gaussian IC-CI to within a bounded gap of one bit, independently of the values of all channel parameters. Using this constant-gap characterization, the generalized degrees of freedom (GDoF) region is determined. It is shown that the introduction of the common message leads to an increase in the GDoF over that achievable over the Gaussian interference channel without a common message, and hence to an unbounded improvement in the achievable rate. 
A surprising feature of the capacity-within-one-bit result is that most of the available benefit (i.e., to within one bit of capacity) due to the common message is achieved through a simple and explicit coding scheme that involves {\em independent} signaling at the two transmitters so that, in effect, this scheme forgoes the opportunity for transmitter cooperation that is inherently available due to shared knowledge of the common message at both transmitters.
\end{abstract}



\newpage

\section{Introduction}

\IEEEPARstart{T}{he} interference channel (IC) consists of two transmit-receive pairs that communicate over a common noisy medium. Each transmitter must convey an individual message to its paired receiver. For even this elemental network, even after decades of research, the capacity region is known only in some special cases \cite{Carleial75, Sato, Gamal_Costa, Annapureddy_Veeravalli, Shang2009, Motahari_Khandani, Benzel}. It is remarkable also that in the general case, the rate region proposed by Han and Kobayashi in 1981 in \cite{Han_Kobayashi} (henceforth, the HK rate region) remains the best known inner bound to date on the capacity region\footnote{In \cite{CMG_IC_achievable_region}, the authors proposed a coding scheme and the corresponding achievable rate, which was shown to be at least as big as the HK rate region. However, in \cite{CMG_Gamal_IC_HK_region}, the region of \cite{CMG_IC_achievable_region} is shown to be equal to the HK rate region. Recently, \cite{hodtani_IC_rate_region} has claimed an improvement over the HK rate region but it is not clear if this improvement is strict.}. The intractability of the exact capacity characterization notwithstanding however, Etkin et al. \cite{Etkin_Tse_Wang} made significant progress by proving -- via a simple and explicit HK scheme (that involves rate-splitting/partial interference decoding) -- that the corresponding rate region is within a universal gap of no more than one bit of the capacity region regardless of the values of the channel parameters. This constant gap result was obtained for the scalar Gaussian IC in \cite{Etkin_Tse_Wang}, and was recently generalized, also through explicit inner and outer bounds on the capacity region, by Karmakar and Varanasi in \cite{Constant-Gap-Karmakar-MV-2011} to the multiple-input multiple-output (MIMO) Gaussian IC with an arbitrary number of antennas at each of the four terminals.

\begin{figure}[b]
\vspace{-1cm}
\begin{center}
\includegraphics[bb=30bp 180bp 650bp 420bp,clip,scale=0.6]{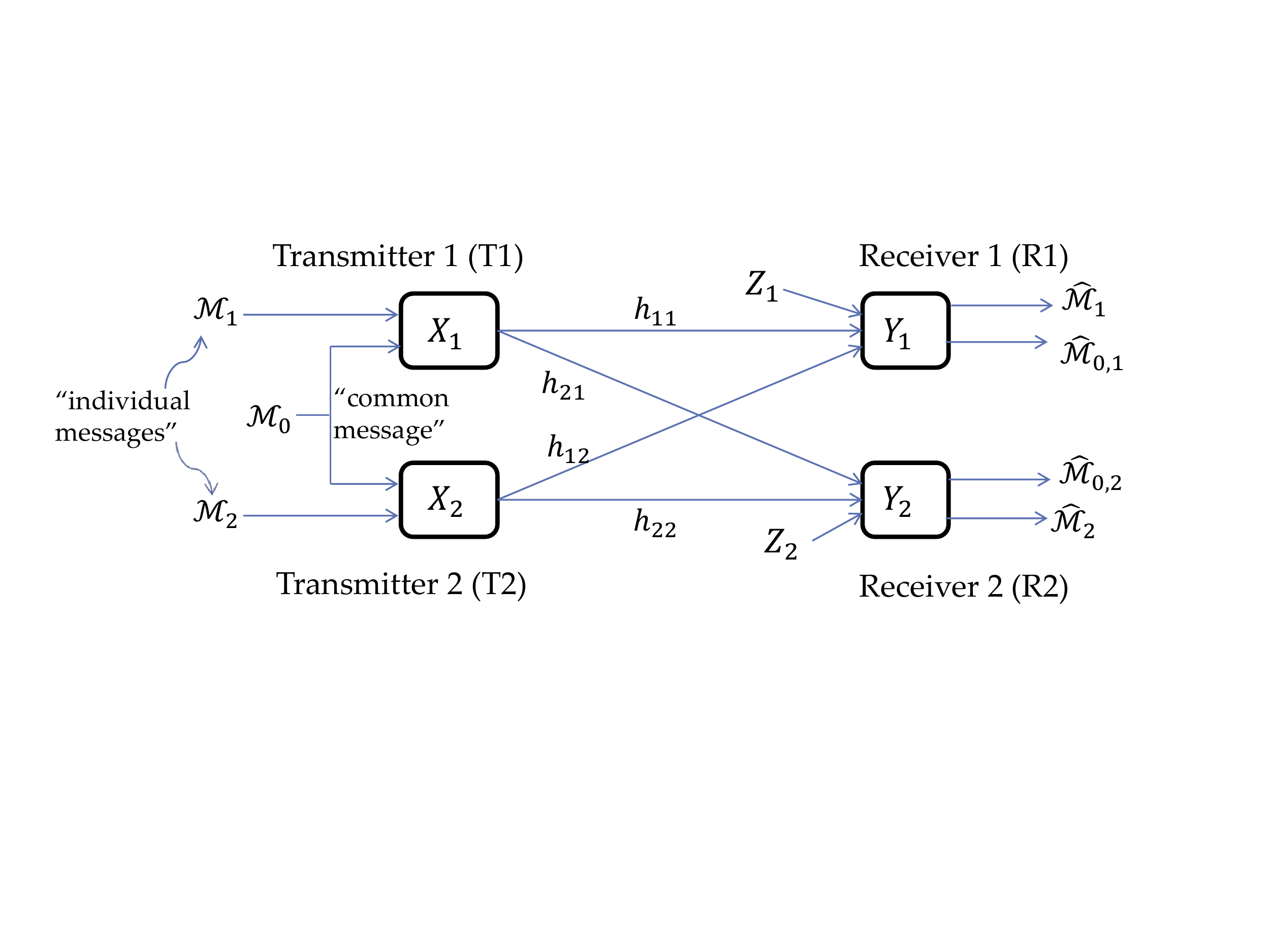} \vspace{-.4cm}
\newline Let $ \snr_i \define |h_{ii}|^2 ~ ~ \mbox{and} ~ ~ \inr_i \define |h_{ij}|^2$. \newline
For the symmetric Gaussian IC-CI, $\snr_1 = \snr_2$ and $\inr_1 = \inr_2$.
\end{center}\vspace{-0.8cm}
\caption{The Gaussian Interference Channel with Common Information (IC-CI)}
\label{fig: gaussian ICCI journal}
\end{figure}

A generalization of the IC, known as the IC with common information (henceforth referred to as the IC-CI), has also been considered in the literature (see Fig. \ref{fig: gaussian ICCI journal}). In the IC-CI, in addition to the two individual messages, both transmitters have a common message, which they must communicate to both receivers\footnote{The term interference channel or IC, henceforth, refers to the case of no common information}. The discrete memoryless (DM) IC-CI was first studied by Tan \cite{tan_IC-CI_rate_region} to obtain inner and outer bounds to the capacity region. More recently, Jiang, Xin, and Garg \cite{jiang_xin_garg_IC_common_info} improved Tan's inner-bound and gave the best-known achievable rate region to date (the rate regions of \cite{Cao_Chen_IC-CI_rate_region} and \cite{jiang_xin_garg_IC_common_info} are equivalent). The coding scheme of \cite{jiang_xin_garg_IC_common_info}, referred henceforth as the JXG scheme, is a generalization of the HK coding scheme. More specifically, the JXG scheme borrows the message-splitting idea from the HK scheme, and involves splitting the individual message at each sender into the public and the private sub-messages, where the former is to be decoded by both receivers while the latter is intended only for the paired receiver. Further, these (sub-)messages are encoded using a three-level superposition encoding scheme with the order of superposition being the common message, followed by the public sub-message, and finally the private sub-massage at the top-most level. Finally, the receivers employ joint typical decoding to extract the desired messages. The main idea of the JXG scheme is to have an identical codeword for the common message at both transmitters, which allows the transmit signals to be {\em dependent}, thus provisioning for a {\em collaborative} transmission of the common message. It is important to  note that the JXG region coincides with the capacity region in some special cases such as the IC-CI with strong interference \cite{maric_yates_kramer_stong_IC-CI}, the deterministic IC-CI, a generalization of the Gamal-Costa deterministic IC \cite{Gamal_Costa}, \cite{jiang_xin_garg_IC_common_info}, and a class of semi-deterministic IC-CI \cite{chong_motani_semideterm_IC-CI}.

While all the above results concern the DM IC-CI, the Gaussian version of this channel has also been explicitly studied. Indeed, the authors of \cite{jiang_xin_garg_IC_common_info} itself propose a class of coding schemes for the real-valued Gaussian IC-CI by specializing their corresponding result on the DM IC-CI. This class of coding schemes is nonspecific in that it consists of uncountably many three-level Gaussian linear superposition coding schemes that are parameterized by four real numbers. As a result, it is impossible to comment on how the rate region of \cite{jiang_xin_garg_IC_common_info}, which is the union of rate regions achieved by each member of that class of coding schemes, relates to the capacity region. Further, for the real Gaussian IC-CI, an outer-bound to the capacity region has been proposed in \cite{cao_chen_IC-CI_outer-bounds}. While this bound is tighter than the one by Tan \cite{tan_IC-CI_rate_region}, nothing is known about its tightness relative to the capacity region. Thus, in summary, the known inner and outer bounds fail to provide any guarantee in general on the closeness of the achievable rates to the capacity region.

\begin{figure}[t]
\centering
\protect{\includegraphics[bb=0bp 0bp 700bp 400bp,clip,scale=0.5]{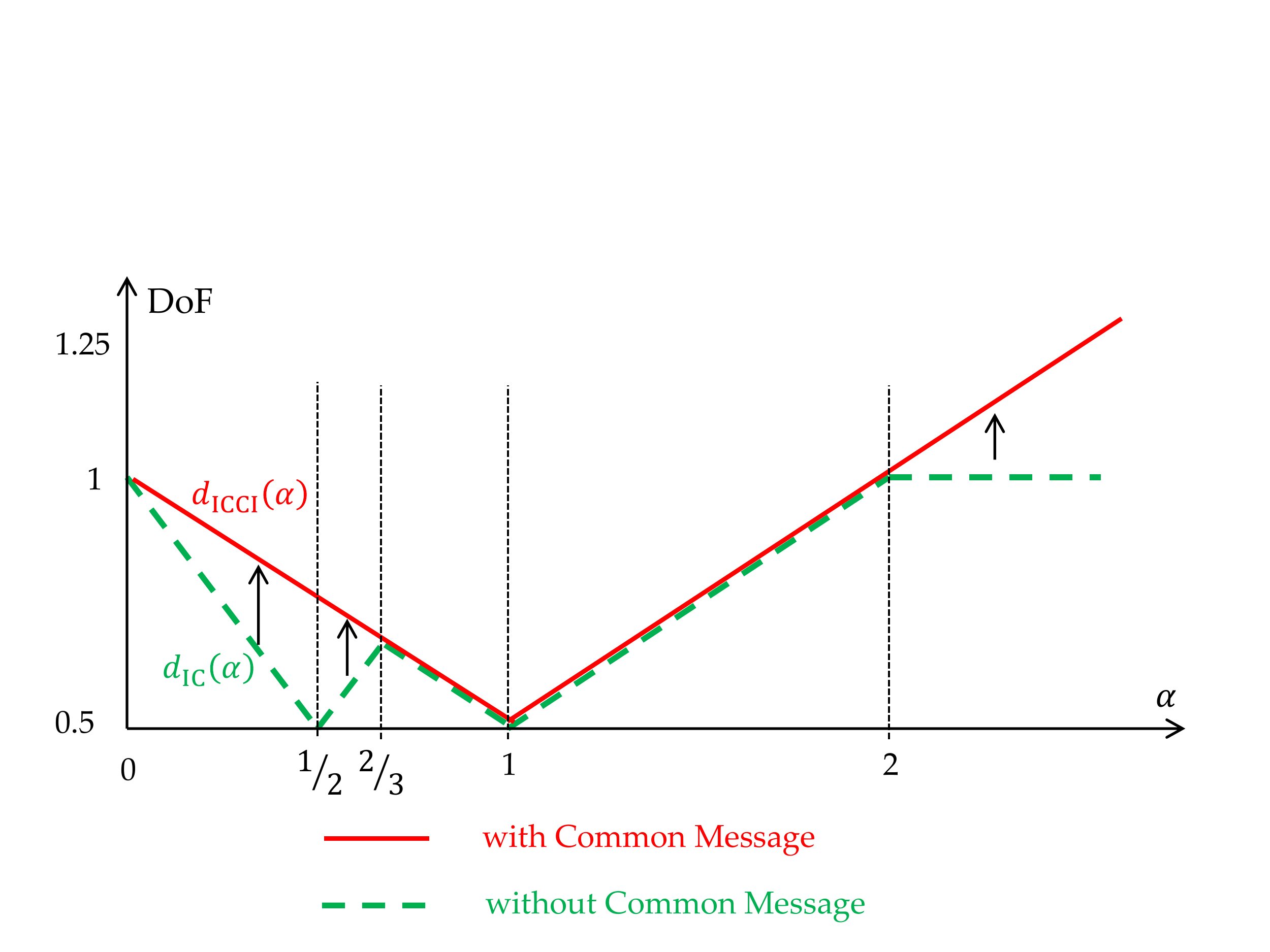}}  \vspace{-0.9cm}
\caption{Maximum Number of Degrees of Freedom Achievable per User over the Symmetric Gaussian IC-CI} \label{fig: GDoF curve symmetric IC-CI SISO}  \vspace{-1cm}
\end{figure}

In this paper, we study the complex-valued Gaussian IC-CI, and determine its capacity region to within a gap of one bit independently of the values of the channel parameters. To establish this result, we first choose an explicit two-level Gaussian superposition coding scheme, which can be seen as being a member of the class of (uncountably many) coding schemes of Jiang et al., and then adopt its achievable rate region as our explicit inner-bound to the capacity region. Next, a new explicit outer-bound to the capacity region is proposed, which has a shape similar to that of the derived inner-bound. This facilitates an easy comparison of the inner and outer-bounds. Further, this comparison reveals that the gap between the two bounds is at most one bit irrespective of the channel parameters, leading to the sought-after characterization of the capacity region of the Gaussian IC-CI to within a bounded gap of one bit.

Using the above constant gap result, we determine the generalized degrees of freedom (GDoF) region, which denotes the rate of growth, in the regime of high nominal $\snr$, of the capacity region with respect to $\log \snr$, when the ratio $\frac{\log \snr_i}{\log \snr} = \alpha_{ii}$ and $\frac{\log \inr_i}{\log \snr} = \alpha_{ij}$ with $i,j \in\{1,2\}$ and $j\neq i$ \cite{Etkin_Tse_Wang}. Focusing on the Gaussian IC-CI where $\alpha_{ii}=1$ and $\alpha_{ij}=\alpha$, so that the channel is symmetric with respect to two transmit-receive pairs, and defining the per-user DoF to be equal to half of the sum GDoF achievable over the channel, we plot in Fig. \ref{fig: GDoF curve symmetric IC-CI SISO} the maximum number of DoF achievable per user with and without the common message. From this figure, it is evident that over a wide range of values of $\alpha$, the presence of common message can significantly enhance the per-user DoF. In other words, an unbounded increase in the achievable rate is possible due to the presence of the common message. This result is interesting because in the cases of the multiple-access and the broadcast channels, the common message can not improve the DoF. Indeed, our result is the first instance in which a common message provides a DoF enhancement. An heuristic explanation of this phenomenon is that in the absence of the common message, the total DoF available at the receivers can not be utilized, whereas, on the contrary, in the presence of the common message, these unused DoF can be completely exploited to produce a DoF improvement.

Based on the forgoing discussion, it would be natural to expect that the constant gap to capacity (and GDoF-region-optimal) coding scheme would involve sending correlated signals from the two transmitters, and hence that some form of transmitter cooperation would in fact be the {\em key} to realizing the potentially unbounded rate improvement promised due to the common message. Remarkably however, the coding scheme proposed here employs independent signaling across the transmitters, as in the case of the IC -- where correlated signaling is not an option, of course-- and yet, it is achieves the capacity region to within one bit. Because of the independence between the two inputs, this scheme involves only a simple and explicit two-level Gaussian superposition coding at each sender (as opposed to the nonspecific and typical three-level JXG superposition schemes). The key insight that emerges due to this paper is therefore that independent inputs, and hence non-cooperative transmission, is optimal to within one bit of the capacity region of the Gaussian IC-CI.

The rest of the paper is organized as follows. The next section describes the model of the Gaussian IC-CI and states some important definitions. The subsequent section contains a brief overview of the coding scheme proposed by Jiang et al. \cite{jiang_xin_garg_IC_common_info}. In Section \ref{subsec: main results ICCI journal}, we describe our main results regarding the capacity region and the GDoF region. In Section \ref{subsec: benefit of M0 ICCI journal}, we show that the common message can enhance the DoF. In Section \ref{subsec: intuition ICCI journal}, we explain how the DoF benefit can be realized with just independent inputs through a simple example. The proofs of the main results are given in Sections \ref{app: proof of thm: outer-bound IC-CI journal}-\ref{app: proof of thm: per-user DoF symmetric IC-CI journal}. Finally, the paper concludes with Section \ref{sec: conclusion ICCI journal}.


\section{Channel Model of the Gaussian IC-CI and Definitions}

We begin by formally defining the Gaussian IC-CI model and then provide some key definitions, in terms of which the main results of this paper are conveniently stated.

\subsection{The Gaussian IC-CI}
The IC-CI (shown in Fig. \ref{fig: gaussian ICCI journal}) consists of two transmitters, T1 and T2, and their corresponding receivers, R1 and R2, and has three messages that need to be conveyed to the receivers. In particular, each transmitter needs to communicate an individual message to its paired receiver, and additionally, both transmitters together need to convey a common message to both receivers. The input-output relationship for the Gaussian IC-CI is described by the following two equations:
\begin{eqnarray*}
Y_1(t)  & = & h_{11} X_1(t) + h_{12} X_2(t) + Z_1(t) \\
Y_2(t)  & = & h_{21} X_1(t) + h_{22} X_2(t) + Z_2(t),
\end{eqnarray*}
where at time $t$, $Y_1(t) \in \mathbb{C} $ and $Y_2(t)\in \mathbb{C} $ are respectively the signals received by R1 and R2; $X_1(t)\in \mathbb{C} $ and $X_2(t) \in \mathbb{C}$ are the signals transmitted by T1 and T2, respectively; $Z_1(t), ~ Z_2(t) \sim \mathcal{C}\mathcal{N}(0,1)$ are the additive Gaussian\footnote{Here, $\mathcal{C}\mathcal{N}(0,\sigma^2)$ denotes a circularly symmetric complex Gaussian distribution with mean $0$ and variance $\sigma^2$.} noises at the two receivers and the noise realizations are assumed to be independent and identically distributed (i.i.d.) across time; $h_{ij} \in \mathbb{C}$ represents the channel fading coefficient between the $i^{th}$ receiver and the $j^{th}$ transmitter; and we impose a power constraint of $1$ on the transmit signals. Finally, the channel coefficients are taken to be deterministic and known to all terminals.

The signal-to-noise ratio (SNR) and the interference-to-noise ratio (INR) corresponding to the two receivers are defined as follows: For each $i \in \{1,2\}$,
\[
\snr_i \define |h_{ii}|^2 ~ ~ \mbox{ and } ~ ~ \inr_i \define |h_{ij}|^2
\]
with $j \in \{1,2\}$ such that $j \not= i$.


{\em Notation: } For any vector $V(t)$ and an $n \in \mathbb{N}$, $V^n$ is defined as the vector with entries $V(1)$, $V(2)$, $\cdots$, $V(n)$. For instance,
\[
Y_i^n = \begin{bmatrix} Y_i^*(1) & Y_i^*(2) & \cdots & Y_i^*(n) \end{bmatrix}^*
\]
Further, $H \define (h_{11},h_{12},h_{21},h_{22})$.

Consider now an $(n,R_0,R_1,R_2,P_e^{(n)})$ coding scheme for the Gaussian IC-CI. It consists of the following components:
\begin{itemize}
\item messages ${\cal M}_0$, ${\cal M}_1$, and ${\cal M}_2$, where ${\cal M}_1$ and ${\cal M}_2$ are the two individual messages, ${\cal M}_0$ is the common message, all messages are independent, and ${\cal M}_i$ is distributed uniformly over the set $\Big\{ 1,2,3,\cdots,2^{nR_i}\Big\}$ of cardinality $2^{nR_i}$;
\item encoding functions $f_1(\cdot)$ and $f_2(\cdot)$, which are used by the transmitters to generate the transmit signal so that for each $i \in \{1,2\}$,
    \[
    X_i(t) = f_i\Big( {\cal M}_0, {\cal M}_i, t, H\Big) \mbox{ and } \frac{1}{n} \sum_{t=1}^n \mathbb{E} |X_i(t)|^2 \leq 1;
    \]
\item decoding functions $g_1(\cdot)$ and $g_2(\cdot)$, which are used by the two receivers to compute the estimates of their desired messages so that
    \[
    \hat{M}_{0i},\hat{M}_i = g_i \Big(Y_i^n, H\Big);
    \]
    and
\item probability of error $P_e^{(n)}$, which is defined as
    \[
    P_e^{(n)} \define \mathrm{Pr} \Big\{ {\cal M}_1 \not= \hat{M}_1 \mbox{ or } {\cal M}_2 \not= \hat{M}_2 \mbox{ or } {\cal M}_0 \not= \hat{M}_{01} \mbox{ or } {\cal M}_0 \not= \hat{M}_{02} \Big\}.
    \]
\end{itemize}
The achievability of the rate $3$-tuple $(R_0,R_1,R_2)$ is defined as follows.
\begin{definition}[Achievability of the rate $3$-tuple]
A rate $3$-tuple $(R_0,R_1,R_2)$ is said to be achievable if there exists a sequence of $(n,R_0,R_1,R_2,P_e^{(n)})$ coding schemes such that $P_e^{(n)} \to 0$ as $n \to \infty$.
\end{definition}

The capacity region and the GDoF region are defined as follows.
\begin{definition}[Capacity region]
The capacity region $\mathbf{C}(H)$ is defined as the closure of the set of all achievable rate $3$-tuples.
\end{definition}

Suppose $\mathbb{R}_+^b$ denotes the set of all $b$-tuples of the non-negative real numbers.
\begin{definition}[GDoF region]
For a vector $\overline{\mathbf{\alpha}} = (\alpha_{11}, \alpha_{12},\alpha_{21},\alpha_{22})^T \in \mathbb{R}_+^4$, the GDoF region $\mathbf{D}(\overline{\alpha})$ is defined as
\begin{eqnarray*}
\lefteqn{ \mathbf{D}(\overline{\mathbf{\alpha}}) = \Biggl\{ (d_0,d_1,d_2) \in \mathbb{R}_+^3 \biggl| ~\mbox{for a } P > 0, ~  |h_{ij}|^2 = P^{\alpha_{ij}} ~\! \forall ~i,j\in\{1,2\} \biggr. \Biggr.  } \\
&& {} \hspace{2cm} \Biggl. \mbox{ and } (R_0,R_1,R_2) \in \mathbf{C}(H) \mbox{ such that } d_k = \lim_{P \to \infty} \frac{R_k}{\log_2 P} ~\! \forall k = 1,2,3 \Biggr\}.
\end{eqnarray*}
\end{definition}

\subsection{More Definitions}

As mentioned in the introduction, we make use of the rate region of Jiang et al. \cite{jiang_xin_garg_IC_common_info} to derive an inner-bound to the capacity region ${\bf C}(H)$. In order to state their rate region, which is done in Theorem \ref{thm: JXG rate region DM IC-CI journal} of the next section, we need the following three definitions.

Consider some jointly distributed random variables $(U_0,U_1,U_2,X_1,X_2,Y_1,Y_2)$. A class of their joint distributions is defined below.
\begin{definition} \label{def: calP ICIC journal}
For random variables $(U_0,U_1,U_2,X_1,X_2,Y_1,Y_2)$, the set of their joint probability distributions $p(\cdot)$ that factor as
\[
p(u_0,u_1,u_2,x_1,x_2,y_1,y_2) = p(u_0) p(u_1,x_1|u_0) p(u_2,x_2|u_0) p(y_1,y_2|x_1,x_2)
\]
is denoted by ${\cal P}$.
\end{definition} \label{def: mutual informs ICCI journal}
We now define some mutual information terms involving these random variables.
\begin{definition}
For a $p(\cdot) \in {\cal P}$ and an $i \in \{1,2\}$, if $j \in \{1,2\}$ such that $j \not=i$, then
\begin{eqnarray*}
\begin{array}{ccccccc}
a_i & \define & I(X_i;Y_i|U_0,U_i,U_j),  &   & d_i & \define & I(X_i;Y_i|U_0,U_j), \\
e_i & \define & I(X_i, U_j;Y_i|U_0,U_i), &   & g_i & \define & I(X_i, U_j;Y_i|U_0), \\
& & & g_i' \define I(U_0,X_i,U_j;Y_i). & & &
\end{array}
\end{eqnarray*}
\end{definition}
Using these mutual information terms, we define a subset of $\mathbb{R}_+^3$ as follows.
\begin{definition} \label{def: fixed JXG region ICCI journal}
For a $p \in \mathcal{P}$, the region ${\cal R}(p) \subset \mathbb{R}_+^3$ is defined as the set of all $3$-tuple $(R_0,R_1,R_2)$ that satisfy the following constraints: $R_0,R_1,R_2 \geq 0$ and
\begin{eqnarray*}
\begin{array}{ccccccc}
R_0 + R_1 & \leq & g_1', & \hspace{-1cm} \hspace{-1cm}   &  R_0 + R_2 & \leq & g_2', \\
R_1 & \leq & d_1, & \hspace{-1cm} \hspace{-1cm}&  R_2 & \leq & d_2, \\
& & & \hspace{-1cm}R_1 + R_2 \leq e_1 + e_2,\hspace{-1cm} & & & \\
R_1 + R_2 & \leq & a_1 + g_2, &\hspace{-1cm}\hspace{-1cm} & R_1 + R_2 & \leq & a_2 + g_1, \\
R_0 + R_1+R_2 & \leq & a_1 + g_2', &\hspace{-1cm}\hspace{-1cm} & R_0 + R_1 + R_2 & \leq & a_2 + g_1', \\
2R_1 + R_2 & \leq & a_1 + g_1 + e_2, &\hspace{-1cm}\hspace{-1cm} & R_1 + 2R_2 & \leq & a_2 + g_2 + e_1, \\
R_0 + 2R_1 + R_2 & \leq & a_1 + g_1' + e_2, &\hspace{-1cm}\hspace{-1cm} & R_0 + R_1 + 2R_2 & \leq & a_2 + g_2' + e_1.
\end{array}
\end{eqnarray*}
\end{definition}

We now define two subsets of $\mathbb{R}_+^3$, namely, $\mathbf{C}_{\rm inner}(H)$ and $\mathbf{C}_{\rm outer}(H)$, which are later proved in Theorems \ref{thm: inner-bound IC-CI journal} and \ref{thm: outer-bound IC-CI journal} to be the inner and outer bounds to ${\bf C}(H)$, respectively. These regions are given in Definitions \ref{def: Cinner ICCI journal} and \ref{def: Couter ICCI journal}, respectively, using some functions of $H$, which are defined next.
\begin{definition} \label{def: numbers of Cinner ICCI journal}
Suppose $\mathbf{\mathsf{C}}(P) \define \log_2 (1+P)$ for a $P \geq 0$. Then
\begin{eqnarray*}
x_{21} & \define & \min\left( 1,\frac{1}{|h_{21}|^2} \right), \hspace{-7pt} \hspace{5cm} x_{12} \define \min \left( 1,\frac{1}{|h_{12}|^2} \right), \\
A_1 & \define & \Cpp \left( \frac{|h_{11}|^2 x_{21}}{1+|h_{12}|^2 x_{12}} \right), \hspace{-7pt} \hspace{4.5cm}  A_2 \define \Cpp \left( \frac{|h_{22}|^2 x_{12}}{1+|h_{21}|^2 x_{21}} \right), \\
D_1 & \define & \Cpp \left( \frac{|h_{11}|^2}{1+|h_{12}|^2 x_{12}} \right), \hspace{-7pt} \hspace{4.5cm} D_2 \define \Cpp \left( \frac{|h_{22}|^2}{1+|h_{21}|^2 x_{21}} \right), \\
E_1 & \define & \Cpp \left( \frac{|h_{11}|^2 x_{21} + |h_{12}|^2 (1-x_{12})}{1+|h_{12}|^2 x_{12}} \right), \hspace{2cm} \hspace{-3pt}  E_2 \define \Cpp \left( \frac{|h_{22}|^2 x_{12} + |h_{21}|^2 (1-x_{21})}{1+|h_{21}|^2 x_{21}} \right), \\
G_1 & \define & \Cpp \left( \frac{|h_{11}|^2 + |h_{12}|^2(1-x_{12})}{1+|h_{12}|^2 x_{12}} \right), \hspace{2.5cm} G_2 \define \Cpp \left( \frac{|h_{22}|^2 + |h_{21}|^2(1-x_{21})}{1+|h_{21}|^2 x_{21}} \right), \\
G_1' & \define & \Cpp \left( \frac{1+|h_{11}|^2+|h_{12}|^2}{1+|h_{12}|^2 x_{12}} -1 \right),  \hspace{2.7cm} \hspace{-2pt} G_2' \define \Cpp \left( \frac{1+|h_{22}|^2+|h_{21}|^2}{1+|h_{21}|^2 x_{21}} -1 \right).
\end{eqnarray*}
\end{definition}
Consider some more non-negative real-valued functions of $H$, which are required for defining $\mathbf{C}_{\rm outer}(H)$.
\begin{definition} \label{def: numbers Couter ICCI journal}
Consider the following parameters:
\begin{eqnarray*}
\overline{A}_1 & \define & \Cpp \left( \frac{|h_{11}|^2}{1+|h_{21}|^2} \right), \hspace{-5pt} \hspace{2.5cm}  \overline{A}_2 \define \Cpp \left( \frac{|h_{22}|^2}{1+|h_{12}|^2} \right), \\
\overline{D}_1 & \define & \Cpp \left( |h_{11}|^2 \right), \hspace{-9pt} \hspace{3.7cm} \overline{D}_2 \define \Cpp \left( |h_{22}|^2 \right), \\
\overline{E}_1 & \define & \Cpp \left( |h_{12}|^2+\frac{|h_{11}|^2}{1+|h_{21}|^2} \right), \hspace{1cm} \hspace{-3pt}  \overline{E}_2 \define \Cpp \left( |h_{21}|^2+\frac{|h_{22}|^2}{1+|h_{12}|^2} \right), \\
\overline{G}_1 & \define & \Cpp \left( |h_{11}|^2 + |h_{12}|^2 \right), \hspace{2cm} \overline{G}_2 \define \Cpp \left( |h_{22}|^2 + |h_{21}|^2\right), \\
\overline{G}_1' & \define & \Cpp \left( [|h_{11}|+|h_{12}|]^2\right),  \hspace{2cm} \hspace{-2pt} \overline{G}_2' \define \Cpp \left( [|h_{22}|+|h_{21}|]^2 \right).
\end{eqnarray*}
\end{definition}

We define now the two subsets $\mathbf{C}_{\rm inner}(H)$ and $\mathbf{C}_{\rm outer}(H)$ in terms of these real numbers.
\begin{definition} \label{def: Cinner ICCI journal}
The region $\mathbf{C}_{\rm inner}(H) \subset \mathbb{R}^3_+$ is defined as the set of all $3$-tuple $(R_0,R_1,R_2)$ that satisfy the following constraints: $R_0,R_1,R_2 \geq 0$ and
\begin{eqnarray*}
\begin{array}{ccccccc}
R_0 + R_1 & \leq & G_1', & \hspace{-1.5cm} \hspace{-1.5cm}  &  R_0 + R_2 & \leq & G_2', \\
R_1 & \leq & D_1, & \hspace{-1.5cm} \hspace{-1.5cm} &  R_2 & \leq & D_2, \\
& & & \hspace{-1.5cm} R_1 + R_2 \leq E_1 + E_2, \hspace{-1.5cm} & & & \\
R_1 + R_2 & \leq & A_1 + G_2, &\hspace{-1.5cm} \hspace{-1.5cm}  & R_1 + R_2 & \leq & A_2 + G_1, \\
R_0 + R_1+R_2 & \leq & A_1 + G_2', &\hspace{-1.5cm} \hspace{-1.5cm}  & R_0 + R_1 + R_2 & \leq & A_2 + G_1', \\
2R_1 + R_2 & \leq & A_1 + G_1 + E_2, &\hspace{-1.5cm} \hspace{-1.5cm}  & R_1 + 2R_2 & \leq & A_2 + G_2 + E_1, \\
R_0 + 2R_1 + R_2 & \leq & A_1 + G_1' + E_2, &\hspace{-1.5cm} \hspace{-1.5cm} & R_0 + R_1 + 2R_2 & \leq & A_2 + G_2' + E_1.
\end{array}
\end{eqnarray*}
\end{definition}

\begin{definition} \label{def: Couter ICCI journal}
The region $\mathbf{C}_{\rm outer}(H) \subset \mathbb{R}^3_+$ is defined as the set of all $3$-tuple $(R_0,R_1,R_2)$ that satisfy the following constraints: $R_0,R_1,R_2 \geq 0$ and
\begin{eqnarray*}
\begin{array}{ccccccc}
R_0 + R_1 & \leq & \overline{G}_1', & \hspace{-1.5cm} \hspace{-1.5cm}  &  R_0 + R_2 & \leq & \overline{G}_2', \\
R_1 & \leq & \overline{D}_1, & \hspace{-1.5cm} \hspace{-1.5cm} &  R_2 & \leq & \overline{D}_2, \\
& & & \hspace{-1.5cm} R_1 + R_2 \leq \overline{E}_1 + \overline{E}_2, \hspace{-1.5cm} & & & \\
R_1 + R_2 & \leq & \overline{A}_1 + \overline{G}_2, &\hspace{-1.5cm} \hspace{-1.5cm}  & R_1 + R_2 & \leq & \overline{A}_2 + \overline{G}_1, \\
R_0 + R_1+R_2 & \leq & \overline{A}_1 + \overline{G}_2', &\hspace{-1.5cm} \hspace{-1.5cm}  & R_0 + R_1 + R_2 & \leq & \overline{A}_2 + \overline{G}_1', \\
2R_1 + R_2 & \leq & \overline{A}_1 + \overline{G}_1 + \overline{E}_2, &\hspace{-1.5cm} \hspace{-1.5cm}  & R_1 + 2R_2 & \leq & \overline{A}_2 + \overline{G}_2 + \overline{E}_1, \\
R_0 + 2R_1 + R_2 & \leq & \overline{A}_1 + \overline{G}_1' + \overline{E}_2, &\hspace{-1.5cm} \hspace{-1.5cm} & R_0 + R_1 + 2R_2 & \leq & \overline{A}_2 + \overline{G}_2' + \overline{E}_1.
\end{array}
\end{eqnarray*}
\end{definition}

We need to define a region $\mathcal{D}(\overline{\alpha}) \subset \mathbb{R}^3_+$ as a function of $\overline{\alpha} = (\alpha_{11},\alpha_{12},\alpha_{21},\alpha_{22})^T$; this region is shown in Theorem \ref{thm: GDoF region ICCI journal} to be equal to the GDoF region ${\bf D}(\overline{\alpha})$. As before, this region $\mathcal{D}(\overline{\alpha})$ is defined in terms of some non-negative real numbers, which are first given below, and subsequently, the region $\mathcal{D}(\overline{\alpha})$ is given by Definition \ref{def: DoF ICCI journal}.

\begin{definition} \label{def: numbers for DoF ICCI journal}
Suppose for two real numbers $a$ and $b$, $(a-b)^+ \define \max\{0,a-b\}$. Then
\begin{eqnarray*}
\begin{matrix}
{\sf a_1} & \define & \left( \alpha_{11} - \alpha_{21} \right)^+, & {\sf a_2} & \define & \left( \alpha_{22} - \alpha_{12} \right)^+, \\[0.3em]
{\sf d_1} & \define & \alpha_{11}, & {\sf d_2} & \define & \alpha_{22}, \\[0.3em]
{\sf e_1} & \define & \max \left\{ \left( \alpha_{11} - \alpha_{21} \right), \alpha_{12} \right\}, & {\sf e_2} & \define & \max \left\{ \left( \alpha_{22} - \alpha_{12} \right), \alpha_{21} \right\}, \\[0.3em]
{\sf g_1} & \define & \max \left\{ \alpha_{11},\alpha_{12} \right\}, & {\sf g_2} & \define & \max \left\{ \alpha_{22},\alpha_{21} \right\}.
\end{matrix}
\end{eqnarray*}
\end{definition}

\begin{definition} \label{def: DoF ICCI journal}
The region $\mathcal{D}(\overline{\alpha}) \subset \mathbb{R}^3_+$ is defined as the set of all $3$-tuple $(d_0,d_1,d_2)$ that satisfy the following constraints: $d_0,d_1,d_2 \geq 0$ and
\begin{eqnarray*}
\begin{array}{ccccccc}
d_0 + d_1 & \leq & {\sf g_1}, & \hspace{-1.5cm} \hspace{-1.5cm}  &  d_0 + d_2 & \leq & {\sf g_2}, \\
d_1 & \leq & {\sf d_1}, & \hspace{-1.5cm} \hspace{-1.5cm} &  d_2 & \leq & {\sf d_2}, \\
& & & \hspace{-1.5cm} d_1 + d_2 \leq {\sf e_1} + {\sf e_2}, \hspace{-1.5cm} & & & \\
d_0 + d_1+d_2 & \leq & {\sf a_1} + {\sf g_2}, &\hspace{-1.5cm} \hspace{-1.5cm}  & d_0 + d_1 + d_2 & \leq & {\sf a_2} + {\sf g_1}, \\
d_0 + 2d_1 + d_2 & \leq & {\sf a_1} + {\sf g_1} + {\sf e_2}, &\hspace{-1.5cm} \hspace{-1.5cm} & d_0 + d_1 + 2d_2 & \leq & {\sf a_2} + {\sf g_2} + {\sf e_1}.
\end{array}
\end{eqnarray*}
\end{definition}

Note here the distinction between $d_i$ and ${\sf d_i}$; while the former denotes the DoF corresponding to ${\cal M}_i$,  the latter is a function of $\overline{\alpha}$ given by Definition \ref{def: numbers for DoF ICCI journal}.

Finally, to state the capacity region of the Gaussian IC-CI to within a constant gap, we need the following definition.
\begin{definition} \label{def: approx ICCI journal}
Let ${\cal R}_1$, ${\cal R}_2$ $\in \mathbb{R}_+^3$. The region ${\cal R}_1$ is said to be within $b\geq0$ bits of the region ${\cal R}_2$ if for any $3$-tuple $(R_0,R_1,R_2) \in {\cal R}_2$, there exists a $3$-tuple $(R_0',R_1',R_2') \in {\cal R}_1$ such that $R_i - R_i' \leq b$ $\forall$ $i\in\{1,2,3\}$.
\end{definition}

\section{An Achievable Rate Region for the DM IC-CI}

As mentioned before, the HK rate region \cite{Han_Kobayashi} is the best-known inner-bound to the capacity region of the general DM IC. Generalizing the coding scheme developed by Han and Kobayashi \cite{Han_Kobayashi}, Jiang, Xin and Garg \cite{jiang_xin_garg_IC_common_info} proposed an achievable rate region for the DM IC-CI (referred to henceforth as the JXG rate region). We state below this rate region and briefly explain the main idea behind the coding scheme of \cite{jiang_xin_garg_IC_common_info} (referred to as the JXG coding scheme), because these ideas are useful when dealing with the Gaussian IC-CI.

We first define the DM IC-CI, which, like the Gaussian IC-CI, consists of two individual messages ${\cal M}_1$ and ${\cal M}_2$, and a common message ${\cal M}_0$. The DM IC-CI is described in terms of the $5$-tuple
\[
\Big({\cal X}_1, {\cal X}_2, {\cal Y}_1, {\cal Y}_2, p(y_1,y_2|x_1,x_2) \Big) ,
\]
where ${\cal X}_1$, ${\cal X}_2$, ${\cal Y}_1$, and ${\cal Y}_2$ are finite-cardinality sets, and the transmit-receive signals at time $t$, namely, $X_1(t)$, $X_2(t)$, $Y_1(t)$, and $Y_2(t)$ belong to sets ${\cal X}_1$, ${\cal X}_2$, ${\cal Y}_1$, and ${\cal Y}_2$, respectively; $p(y_1,y_2|x_1,x_2)$ denotes the conditional transition probability and the channel is memoryless in a sense that
\[
p\left( y_1^n,y_2^n \left| x_1^n, x_2^n \right. \right) = \prod_{t=1}^n p \left( y_1(t), y_2(t) \left| x_1(t),x_2(t) \right. \right).
\]
For the DM IC-CI, the achievability of the rate tuple $(R_0,R_1,R_2)$ and the capacity region ${\bf C}\Big(p(y_1,y_2|x_1,x_2)\Big) $ are defined in a manner similar to their definitions for the Gaussian IC-CI. Then the JXG achievable rate region for the DM IC-CI is given by the following theorem, which is stated using Definitions \ref{def: calP ICIC journal}-\ref{def: fixed JXG region ICCI journal}.

\begin{theorem}[JXG rate region, \cite{jiang_xin_garg_IC_common_info}] \label{thm: JXG rate region DM IC-CI journal}
The region
\[
{\cal R}_{\rm JXG} \define \bigcup_{p \in \mathcal{P}} {\cal R}(p)
\]
is achievable over the DM IC-CI, i.e., $ {\cal R}_{\rm JXG} \subseteq {\bf C}\Big(p(y_1,y_2|x_1,x_2)\Big) $.
\end{theorem}

We now provide a brief explanation of the JXG coding scheme.

\begin{remark}[JXG coding scheme \cite{jiang_xin_garg_IC_common_info}] \label{rem: JXG scheme journal ICCI}
This scheme can be considered as a generalization of the HK coding scheme \cite{Han_Kobayashi} in the sense that it superposes an HK scheme on the common message code. As in the HK scheme, each individual message ${\cal M}_i$ is split into {\em private} and {\em public} sub-messages, denoted as ${\cal M}_{i,pr}$ and ${\cal M}_{i,pu}$. The private sub-message ${\cal M}_{i,pr}$ is to be decoded only by the $i^{th}$ receiver, while the public sub-message ${\cal M}_{i,pu}$ is to be decoded by both the receivers. The transmitters use superposition encoding and  the receivers use joint typical decoding. Furthermore, the rate $3$-tuple $(R_0,R_1,R_2)$ is attained by achieving rates $R_0$, $R_{1,pu}$, $R_{1,pr}$, $R_{2,pu}$, and $R_{2,pr}$ for messages ${\cal M}_0$, ${\cal M}_{1,pu}$, ${\cal M}_{1,pr}$, ${\cal M}_{2,pu}$, and ${\cal M}_{2,pr}$, respectively, so that $R_1 = R_{1,pu} + R_{1,pr}$ and $R_2 = R_{2,pu} + R_{2,pr}$.

The codebooks in the JXG scheme are generated as follows. The messages ${\cal M}_{i,pu}$ and ${\cal M}_{i,pr}$ are taken to be uniformly distributed over sets $\Big\{1, 2, 3, \cdots, 2^{nR_{i,pu}}\Big\}$ and $\Big\{1, 2, 3, \cdots, 2^{nR_{i,pr}}\Big\}$, respectively. A code consisting of $2^{nR_0}$ $n$-length i.i.d. codewords $\mathbf{u}_0(k)$ with $k \in \Big\{1,2,\cdots,2^{nR_0}\Big\}$ is generated (corresponding to $U_0$) according to the probability law $\prod_{t=1}^n p \big( u_0(t) \big)$. Next, for each $i \in \{1,2\}$ and for each codeword $\mathbf{u}_0(k)$, $2^{nR_{i,pu}}$ i.i.d. codewords $\mathbf{u}_i(k,l_i)$ with $l_i \in \Big\{1,2,\cdots, 2^{nR_{i,pu}}\Big\}$ are generated according to $\prod_{t=1}^n p \big( u_i(t) \big| u_0(t) \big)$. Finally, for every $i \in \{1,2\}$ and for each pair of codewords $\big( {\bf u}_0(k), {\bf u}_i(l_i) \big)$, $2^{nR_{i,pr}}$ i.i.d. codewords ${\bf x}_i(k,l_i,m_i)$ with $m_i \in \Big\{1,2,\cdots,2^{nR_{i,pr}}\Big\}$ are generated according to the law $\prod_{t=1}^n p(x_i(t)|u_0(t),u_i(t))$. These codebooks are then revealed to all terminals before the start of data communication.

Now, to encode messages ${\cal M}_0 = m_0$, ${\cal M}_{i,pu} = m_{i,pu}$, and ${\cal M}_{i,pr} = m_{i,pr}$, the $i^{th}$ transmitter transmits codeword ${\bf x}_i \big( m_0, m_{i,pu}, m_{i,pr}\big)$. On the other hand, the receivers compute the estimates of the transmitted common message, two  public sub-messages, and the intended private sub-message by finding unique codewords corresponding to these messages that are jointly typical with its received signal. From these estimates, they decode the desired messages.

Note that in the above encoding scheme, the common message is not conveyed by just the codeword corresponding to $U_0$, but also by those corresponding to $U_1$, $X_1$, $U_2$, and $X_2$. This is because the choices of these codewords depend on the value/realization of the common message. Similarly, the $i^{th}$ public message is encoded not just via $U_i$ but also via $X_i$.
\end{remark}

\section{Main Results and DoF Improvement due to the Common Message}

In this section, we describe our main results and also the key insight obtained about the DoF improvement possible with the common message.

\subsection{The Capacity Region to within One Bit and the GDoF region} \label{subsec: main results ICCI journal}

Here, we provide a constant-gap characterization of the capacity region and then the GDoF region. Toward this end, we first derive an inner-bound to the capacity region ${\bf C}(H)$ in the following theorem, which is stated using Definitions \ref{def: numbers of Cinner ICCI journal} and \ref{def: Cinner ICCI journal}.

\begin{theorem} \label{thm: inner-bound IC-CI journal}
The region $\mathbf{C}_{\rm inner}(H)$ is achievable over the Gaussian IC-CI, i.e.,
\[
\mathbf{C}_{\rm inner}(H) \subseteq \mathbf{C}(H).
\]
\end{theorem}
\begin{IEEEproof}
We use the JXG coding scheme with the following choice for the distribution of the transmit-side random variables: For $i,j \in \{1,2\}$ with $j \not= i$,
\begin{eqnarray*}
U_0 & = &  0, \quad U_i  \sim  \mathcal{C}\mathcal{N}\left( 0,1-x_{ji} \right) \\
X_i & = & U_i + U_{i,pr}, \mbox{ where } U_{i,pr} \sim \mathcal{C}\mathcal{N} \left( 0, x_{ji} \right),
\end{eqnarray*}
and $U_{1}$, $U_2$, $U_{1,pr}$, $U_{2,pr}$ are independent. Note that in choosing the power split between $U_i$ and $U_{i,pr}$, i.e., between the public and the private sub-messages, we have used the key insight from Etkin et. al. in \cite{Etkin_Tse_Wang} for the IC that the private sub-message should appear at the noise floor of the unintended receiver. Further, this choice of $p(u_0,u_1,u_2,x_1,x_2)$ induces a joint distribution on random variables $(U_0,U_1,U_2,X_1,X_2,Y_1,Y_2)$, which is referred in the sequel as $p_G$. Note that $p_G \in {\cal P}$ and is hence a valid, if fringe or degenerate, JXG coding scheme (since $U_0=0$). Now, various mutual information terms, defined in Definition \ref{def: mutual informs ICCI journal}, can be easily evaluated for $p = p_G$ to verify that for each $i \in \{1,2\}$,
\[
a_i = A_i, ~ d_i = D_i, ~ e_i = E_i, ~ g_i = G_i, ~ \mbox{and } g_i' = G_i'.
\]
The details are straightforward and omitted here. This implies that ${\cal R}\big( p_G \big) = \mathbf{C}_{\rm inner}(H)$. Hence, the region $\mathbf{C}_{\rm inner}(H)$ is achievable, as per Theorem \ref{thm: JXG rate region DM IC-CI journal}.
\end{IEEEproof}

\begin{remark}[On the coding scheme used in Theorem \ref{thm: inner-bound IC-CI journal}]
The intriguing feature of the apparently degenerate $p_G$ coding scheme used to prove the achievability of $\mathbf{C}_{\rm inner}(H)$ is that the random variable $U_0$, which represents the cooperation between the two transmitters in sending the common message, is set equal to $0$. 
Thus, the transmitters ignore the fact that the common message is known to both of them. Rather, each transmitter combines the common message with its individual message, treats their combination as its new individual message, and then encodes disregarding the fact that a part of this new individual message is also known to the other transmitter. The receivers, on the other hand, must decode the desired messages accounting for the presence of the common message. Remarkably, as proved in Theorem \ref{thm: cap within 1 bit ICCI journal}, this degenerate JXG coding scheme, which in spite of forgoing entirely the opportunity for transmitter cooperation, is capacity optimal to within one bit.
\end{remark}

\begin{remark}[On the JXG coding scheme for the (real) Gaussian IC-CI in \cite{jiang_xin_garg_IC_common_info}] On the contrary, the JXG rate region in \cite[Section VI-A]{jiang_xin_garg_IC_common_info} is (a) specified for the Gaussian IC-CI as a union of a collection of uncountably infinitely many subsets of $\mathbb{R}_+^3$, where the collection is parameterized by four numbers belonging to the set $[0,1]$ and (b) there is no assurance about its distance to the capacity region.
\end{remark}

To show that the inner-bound to the capacity region of Theorem \ref{thm: inner-bound IC-CI journal} is tight up to 1 bit, we need a tight outer-bound. The next theorem gives us such an outer bound, and is stated using Definitions \ref{def: numbers Couter ICCI journal} and \ref{def: Couter ICCI journal}.

\begin{theorem} \label{thm: outer-bound IC-CI journal}
The region $\mathbf{C}_{\rm outer}(H)$ is an outer-bound to the capacity region $\mathbf{C}(H)$ of the Gaussian IC-CI, i.e.,
\[
\mathbf{C}(H) \subseteq  \mathbf{C}_{\rm outer}(H).
\]
\end{theorem}
\begin{IEEEproof}
See Section \ref{app: proof of thm: outer-bound IC-CI journal}.
\end{IEEEproof}

The following result shows that the bounds of Theorems \ref{thm: inner-bound IC-CI journal} and \ref{thm: outer-bound IC-CI journal} are within a one bit gap.

\begin{theorem} \label{thm: cap within 1 bit ICCI journal}
The inner-bound $\mathbf{C}_{\rm inner}(H)$ is within one bit of the outer-bound $\mathbf{C}_{\rm outer}(H)$ $\forall$ $H$. Hence, $\mathbf{C}_{\rm inner}(H)$ is within a bounded gap of one bit to the capacity region $\mathbf{C}(H)$, independently of the channel parameters.
\end{theorem}
\begin{IEEEproof}
See Section \ref{app: proof of thm: cap within 1 bit ICCI journal}.
\end{IEEEproof}


Next, using the previous theorem, the GDoF region of the Gaussian IC-CI is computed.

\begin{theorem} \label{thm: GDoF region ICCI journal}
For the Gaussian IC-CI, the GDoF region is equal to ${\cal D}(\overline{\alpha})$, i.e.,
\[
\mathbf{D}(\overline{\alpha})  = {\cal D}(\overline{\alpha}).
\]
where $ {\cal D}(\overline{\alpha})$ is defined in Definitions  \ref{def: numbers for DoF ICCI journal} and \ref{def: DoF ICCI journal}.
\end{theorem}
\begin{IEEEproof}
Since the inner-bound $\mathbf{C}_{\rm inner}(H)$ is within one bit of the outer-bound $\mathbf{C}_{\rm outer}(H)$ for any $H$, these two regions for tight in the sense of GDoF. Hence, the GDoF region can computed using either one of the two bounds. See Section\ref{app: proof of thm: GDoF region ICCI journal} for the details of this computation.
\end{IEEEproof}

Note that in the special of $d_0 = 0$, i.e., in the absence of the common message, the region ${\cal D}(\overline{\alpha})$ reduces to the well known GDoF region of the IC (without common message) of \cite{Etkin_Tse_Wang}. Moreover, in the special case of $d_1 = d_2 = 0$, i.e., when the individual messages are not transmitted, the above theorem recovers the result on the DoF of the broadcast channel with just a common message.

Having determined the GDoF region, it is now possible to quantify the benefit of having the common message, which is the topic of the next sub-section.

\begin{remark}[On \cite{cao_chen_IC-CI_outer-bounds}]
Four outer-bounds to the capacity region of the Gaussian IC-CI have been proposed before in \cite{cao_chen_IC-CI_outer-bounds}. However, none of these outer-bounds put any constraint on the linear combinations $R_0 + 2R_1 + R_2$ and $R_0 + R_1 + 2R_2$. However, bounds on these linear combinations are important even in the DoF sense, i.e., corresponding to these, we get bounds on $d_0+2d_1+d_2$ and $d_0+d_1+2d_2$ while characterizing the GDoF region, and these bounds can be shown to be non-redundant. Thus, we conjecture that all outer-bounds of \cite{cao_chen_IC-CI_outer-bounds} have an unbounded gap to the capacity region. Hence, their outer-bounds are not used here to obtain a constant-gap characterization of the capacity region.
\end{remark}

\begin{remark}[On \cite{henry_varanasi_ICCI_2011}]
Recently, following the conference version of this paper in \cite{vaze_varanasi_ICCI_isit}, a companion paper by Romero and Varanasi in \cite{henry_varanasi_ICCI_2011} generalizes the inner and outer bounds of Telatar and Tse \cite{Telatar-Tse-IC} for a class of DM semi-deterministic ICs to the class of DM semi-deterministic IC-CIs. When specialized to the Gaussian IC-CI, the result in \cite{henry_varanasi_ICCI_2011} also obtains the one bit gap result. However, as in the case of \cite{Telatar-Tse-IC}, the work in \cite{henry_varanasi_ICCI_2011} obtains inner and outer bounds to the capacity region that are both expressed as the union of a collection of (possibly uncountably) infinitely many subsets of $\mathbb{R}_+^3$, where this collection of  subsets is parameterized by the joint distributions of a certain set of random variables. In other words, the result of \cite{henry_varanasi_ICCI_2011}, while general (it is also applicable, for example, to MIMO IC-CIs), does not provide an explicit characterization of the capacity region, unlike this paper. Another shortcoming of the approach of \cite{henry_varanasi_ICCI_2011}, as in the case with \cite{Telatar-Tse-IC}), is that due to the lack of explicit characterizations of the bounds, it is difficult to obtain further insights, such as proving that the common message can lead to a DoF improvement, as is done in this paper and its conference version \cite{vaze_varanasi_ICCI_isit}.
\end{remark}

\subsection{GDoF Benefit due to Common Message} \label{subsec: benefit of M0 ICCI journal}

To simply quantify the benefit of having a common message, we focus here on the {\em symmetric} Gaussian IC-CI, where $ h_{11} = h_{22}, h_{12} = h_{21},
\alpha_{11} = \alpha_{22} = 1, \alpha_{12} = \alpha_{21} = \alpha, $ and  $ R_1 = R_2 $ and $ d_1 = d_2 $.
Since we achieve an equal number of DoF for the two individual messages in the symmetric case and the sum $(d_0+d_1+d_2)$ represents the total DoF achieved over the channel, the number $\frac{1}{2} (d_0+d_1+d_2)$ represents the per-user DoF. Hence, the quantity
\begin{eqnarray*}
\mathbf{d}_{{\sf IC}} (\alpha) \define \max_{(d_0,d_1,d_2) \in \mathbf{D}(\overline{\mathbf{\alpha}})} \frac{1}{2} \left( d_0 + d_1+d_2 \right), \mbox{ subject to } \left\{ \begin{array}{cc} \alpha_{11} =\alpha_{22} = 1, & d_1 = d_2 \\ \alpha_{12} = \alpha_{21} = \alpha, & d_0 = 0, \end{array} \right\},
\end{eqnarray*}
denotes the maximum number of DoF achievable per user without the common message, whereas
\begin{eqnarray*}
\mathbf{d}_{{\sf ICCI}} (\alpha) \define \max_{(d_0,d_1,d_2) \in \mathbf{D}(\overline{\mathbf{\alpha}})} \frac{1}{2} \left( d_0 + d_1+d_2 \right), \mbox{ subject to } \left\{ \begin{array}{c} \alpha_{11} =\alpha_{22} = 1  \\ \alpha_{12} = \alpha_{21} = \alpha  \end{array}, ~ d_1 = d_2 \right\}. \label{eq: defn d ICCI journal}
\end{eqnarray*}
represents the per user DoF with the common message. As a result, the difference
\[
\mathbf{d}_{\uparrow} (\alpha) \define \mathbf{d}_{{\sf ICCI}} (\alpha) - \mathbf{d}_{{\sf IC}} (\alpha)
\]
signifies the improvement attainable in the per-user DoF due to the common message. Thus, to quantify the benefit of having the common message, it is sufficient to characterize $\mathbf{d}_{\uparrow} (\alpha)$. Toward this end, we first determine $\mathbf{d}_{{\sf IC}}$, and then derive $\mathbf{d}_{{\sf ICCI}}$ to obtain an expression for $\mathbf{d}_{\uparrow} (\alpha)$ in Theorem \ref{thm: per-user DoF symmetric IC-CI journal}.
\begin{theorem}[Etkin et al \cite{Etkin_Tse_Wang}] \label{thm: per-user DoF wo common symmetric IC-CI journal}
Over the symmetric Gaussian IC-CI, we have
\begin{eqnarray}
{\bf d}_{\sf IC}(\alpha) = \begin{cases} 1 - \alpha & \mbox{if } 0 \leq \alpha < \frac{1}{2}, \\
                                           \alpha   & \mbox{if } \frac{1}{2} \leq \alpha < \frac{2}{3}, \\
                                         1 - \frac{\alpha}{2} & \mbox{if } \frac{2}{3} \leq \alpha < 1, \\
                                           \frac{\alpha}{2}   & \mbox{if } 1 \leq \alpha < 2, \\
                                                1             & \mbox{if } 2 \leq \alpha. \end{cases}
\label{eq: thm: per-user DoF wo common symmetric IC-CI journal}
\end{eqnarray}
\end{theorem}
\begin{IEEEproof}
In the absence of the common message or $d_0 = 0$, the IC-CI  reduces to IC, and therefore, we get ${\bf d}_{\sf IC}(\alpha)$ from \cite[equation (25)]{Etkin_Tse_Wang}.
\end{IEEEproof}

\begin{theorem} \label{thm: per-user DoF symmetric IC-CI journal}
Over the symmetric Gaussian IC-CI, we have
\begin{eqnarray}
{\bf d}_{\sf ICCI}(\alpha) =  \begin{cases}
{\bf d}_{\sf IC}(\alpha)+ \frac{\alpha}{2}        & \mbox{if } 0 \leq \alpha < \frac{1}{2}, \\
{\bf d}_{\sf IC}(\alpha) +\frac{2-3\alpha}{2}     & \mbox{if } \frac{1}{2} \leq \alpha < \frac{2}{3}, \\
{\bf d}_{\sf IC}(\alpha)                          & \mbox{if } \frac{2}{3} \leq \alpha < 1, \\
{\bf d}_{\sf IC}(\alpha)                          & \mbox{if } 1 \leq \alpha < 2, \\
{\bf d}_{\sf IC}(\alpha) + \frac{\alpha-2}{2}     & \mbox{if } 2 \leq \alpha,
\end{cases} \label{eq: thm: per-user DoF symmetric IC-CI journal}
\end{eqnarray}
and therefore,
\begin{eqnarray}
{\bf d}_{\uparrow}(\alpha) = \begin{cases}
\frac{\alpha}{2}        & \mbox{if } 0 \leq \alpha < \frac{1}{2}, \\
\frac{2-3\alpha}{2}     & \mbox{if } \frac{1}{2} \leq \alpha < \frac{2}{3}, \\
0                       & \mbox{if } \frac{2}{3} \leq \alpha < 1, \\
0                       & \mbox{if } 1 \leq \alpha < 2, \\
\frac{\alpha-2}{2}      & \mbox{if } 2 \leq \alpha.
\end{cases} \label{eq: increase thm: per-user DoF symmetric IC-CI journal}
\end{eqnarray}
\end{theorem}
\begin{IEEEproof}
See Section \ref{app: proof of thm: per-user DoF symmetric IC-CI journal}.
\end{IEEEproof}
The functions ${\bf d}_{\sf IC}(\alpha)$ and ${\bf d}_{\sf ICCI}(\alpha)$ are plotted in Fig. \ref{fig: GDoF curve symmetric IC-CI SISO}. From this figure, we observe that the common message can lead to a significant improvement in per-user DoF over a wide range of values of $\alpha$.

\begin{remark}
It is interesting that in the cases of the multiple-access and broadcast channels, the introduction of the common message can not result in a DoF improvement. In other words, if either the transmitter or the receivers are cooperating, then at most $\max (1,\alpha)$ DoF can be achieved per user, and these DoF are achievable even in the absence of the common message. Hence, the common message can not produce a DoF improvement over the multiple-access and broadcast channels, unlike in the case of the IC. We provide an example illustrating why ${\bf d}_{\uparrow}(\alpha)$ is non-zero in general for the IC-CI next.
\end{remark}

\subsection{Common Message Leads to a GDoF Improvement: An Illustration} \label{subsec: intuition ICCI journal}

In the absence of a common message, a simple HK coding scheme is constant-gap-to-capacity (and hence GDoF) region optimal for the Gaussian IC \cite{Etkin_Tse_Wang}. In the HK coding scheme, each (individual) message is split into two sub-messages, of which the private sub-message is decoded by just the intended receiver, whereas the public sub-message is to be decoded by both receivers. Thus, each receiver decodes three sub-messages, namely, the intended private and the two public sub-messages, while treating the contribution due to the unintended private sub-message as noise. However, in so doing, all available DoF are not used at the two receivers. In contrast, these unused DoFs are used by the common message in the Gaussian IC-CI to effect a DoF improvement. We next give a concrete example to explicitly illustrate this point.

Consider the symmetric Gaussian IC-CI with $\alpha = 0.6$. While this implies that $|h_{11}|^2 = |h_{22}|^2 = P$ and $|h_{12}|^2 = |h_{21}|^2 = P^{0.6}$ with $P \to \infty$, we assume here for simplicity that $ h_{11} = h_{22} = \sqrt{P} ~ ~ \mbox{ and } ~ ~ h_{21} = h_{12} = \sqrt{P^{0.6}} $.
Further, at $\alpha = 0.6$, $ {\bf d}_{\sf IC}(0.6) = 0.6 $ and $ {\bf d}_{\sf ICCI}(0.6) = 0.7 $ which implies $ {\bf d}_{\uparrow} = 0.1 $.
We will show here that while achieving $0.6$ DoF for the two individual messages, $0.2$ DoF can be achieved for the common message, which implies, in the achievability sense, that ${\bf d}_{\uparrow} = 0.1$. Moreover, we will achieve $0.4$ and $0.2$ DoF for the private and the public sub-messages.

Toward this end, we use a JXG coding scheme with the following choice for the transmit-side random variables: for $i \in \{1,2\}$,
\begin{eqnarray*}
U_0 = 0, \quad  U_i = U_{0i} + U_{i,pu}, \quad \mbox{and} \quad X_i = U_i + U_{i,pr}, \quad \mbox{ where} \\
U_{0i} \sim \mathcal{C}\mathcal{N} \left( 0,P^{-0.2} \right), \quad U_{i,pu} \sim \mathcal{C}\mathcal{N}\left( 0,1 \right), \quad U_{i,pr} \sim \mathcal{C}\mathcal{N} \left( 0, P^{-0.6} \right),
\end{eqnarray*}
and all random variables, namely, $U_{01}$, $U_{1,pu}$, $U_{1,pr}$, $U_{02}$, $U_{2,pu}$, and $U_{1,pr}$ are independent. Here, random variables $\left\{ U_{01}, U_{02} \right\}$, $U_{i,pu}$, and $U_{i,pr}$ carry messages ${\cal M}_0$, ${\cal M}_{i,pu}$, and ${\cal M}_{i,pr}$, respectively. Essentially, under this scheme, the $i^{th}$ transmitter generates independent Gaussian codebooks for random variables $U_{0i}$, $U_{i,pu}$, and $U_{i,pr}$; appropriate codewords of these codebooks are selected in order to encode messages ${\cal M}_0$, ${\cal M}_{i,pu}$, and ${\cal M}_{i,pr}$, respectively; and finally, $X_i$, the transmit signal, is generated as the sum of all three codewords. Since $U_i$ is set equal to the sum of the codewords for $U_{0i}$ and $U_{i,pu}$, this scheme treats the combination of the common message and public sub-message as the new/effective public sub-message.

A minor detail about the above scheme is that it does not strictly satisfy the power constraint but this issue can be easily handled (we omit it to keep the focus on the main point). Moreover, except for this issue, the scheme here is identical to the one suggested in Theorem \ref{thm: inner-bound IC-CI journal}, which achieves the capacity region to within $1$ bit. Also, we have suppressed the time index here, since the signals are i.i.d. across time.

\begin{figure}[t]
\centering
\includegraphics[bb=20bp 20bp 710bp 290bp,clip,scale=0.5]{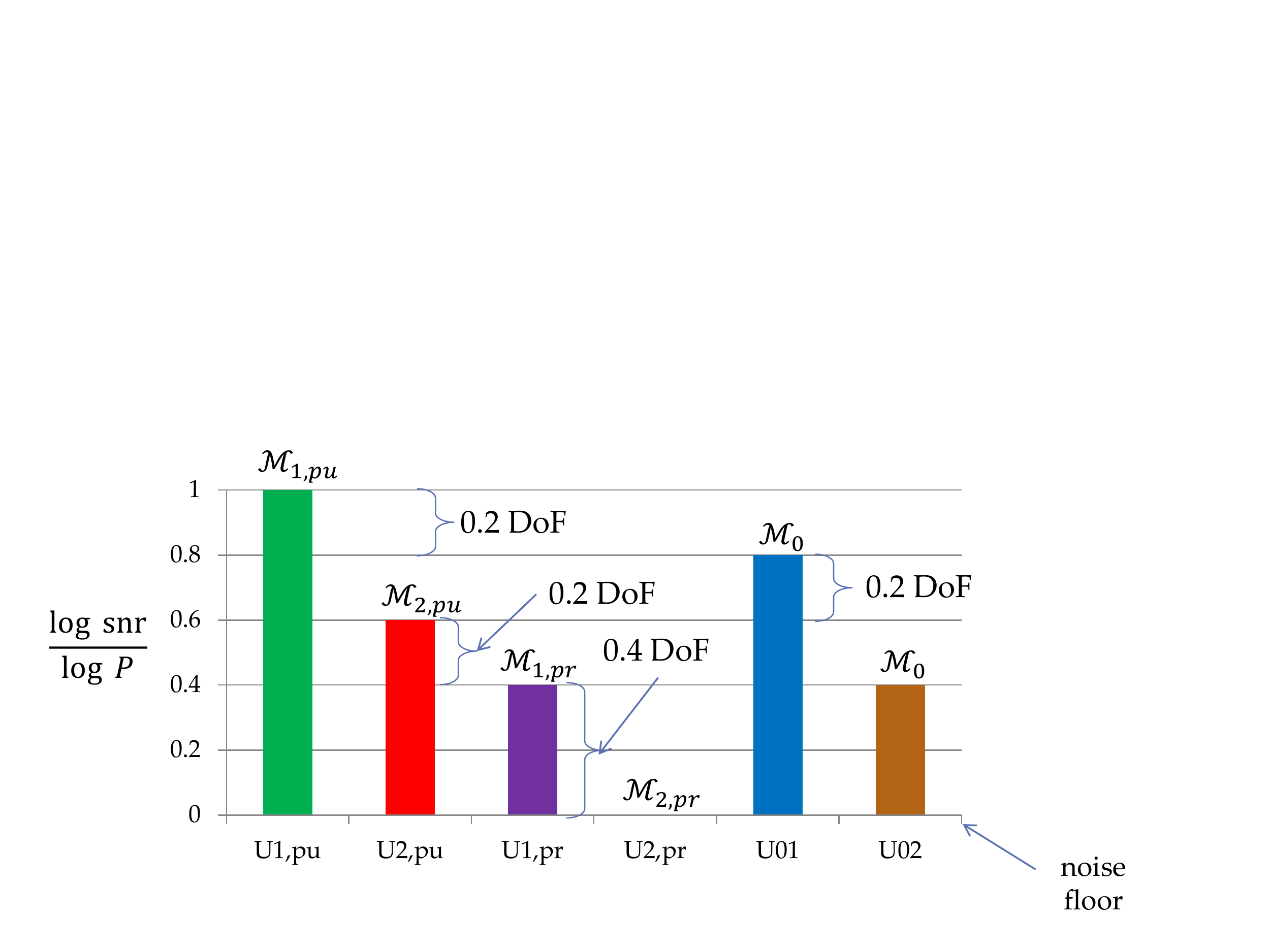} \vspace{-0.75cm}
\caption{Receive Signal Level Space of R1: in presence of ${\cal M}_0$} \vspace{5mm}
\label{fig: exmple scheme with M0 ICCI journal}
\includegraphics[bb=20bp 20bp 710bp 290bp,clip,scale=0.5]{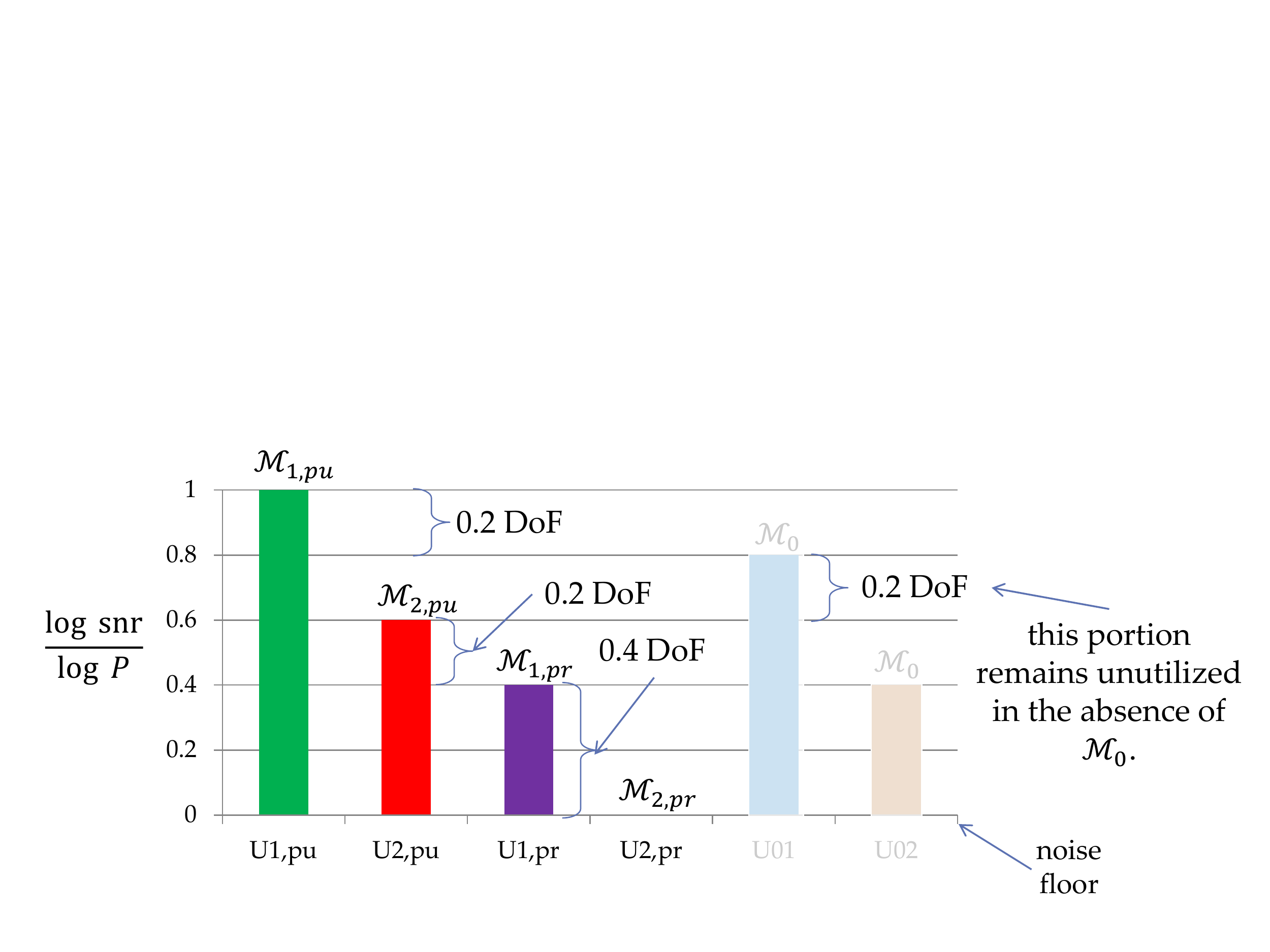}\vspace{-0.75cm}
\caption{Receive Signal Level Space of R1: in absence of ${\cal M}_0$}
\label{fig: exmple scheme no M0 ICCI journal}
\end{figure}

Consider the decoding operation at R1 without loss of generality (due to symmetry). Its received signal can be written as
\begin{eqnarray*}
Y_1 =
\underbrace{\sqrt{P}  ~\! U_{01} + \sqrt{P^{0.6}} ~\! U_{02}}_{\mbox{\small $\begin{array}{c} {\cal M}_0 \\ \snrs = P^{0.8} \end{array}$}}  ~\! +
\underbrace{\sqrt{P} ~\!  U_1}_{\mbox{\small $\begin{array}{c} {\cal M}_{1,pu} \\ \snrs = P \end{array}$}} +
\underbrace{\sqrt{P} ~\!  U_{1,pr}}_{\mbox{\small $\begin{array}{c} {\cal M}_{1,pr} \\ \snrs = P^{0.4} \end{array}$}} + \underbrace{\sqrt{P^{0.6}} ~\!  U_2}_{\mbox{\small $\begin{array}{c} {\cal M}_{2,pu} \\ \snrs = P^{0.6} \end{array}$}} +
\underbrace{\sqrt{P^{0.6}} ~\!  U_{2,pr} + Z_1,}_{\mbox{\small $\begin{array}{c} \mbox{interference+noise} \\ \snrs = 2 ~\! \forall ~\! P\end{array}$}}
\end{eqnarray*}
where $\snrs$ denotes the signal-to-noise ratio (SNR) of the corresponding signal. Since the ratio $\frac{\log \snrs}{\log P}$ has relevance in the DoF calculations, we plot this ratio in Fig. \ref{fig: exmple scheme with M0 ICCI journal} for all messages.

The receiver R1 thus sees equivalently a multiple-access channel, where it needs to decode messages ${\cal M}_0$, ${\cal M}_{1,pu}$, ${\cal M}_{2,pu}$, and ${\cal M}_{1,pr}$, while treating the interference due to $M_{2,pr}$ as noise. To this end, the receiver can employ sequential decoding with successive interference cancellation with the order of decoding being ${\cal M}_{1,pu}$ $\rightarrow$ ${\cal M}_0$ $\rightarrow$ ${\cal M}_{2,pu}$ $\rightarrow$ $M_{1,pr}$, and thereby, it can extract $0.2$, $0.2$, $0.2$, and $0.4$ DoF for these messages respectively (see also Fig. \ref{fig: exmple scheme with M0 ICCI journal}). In other words, R1 can achieve $0.4+0.2=0.6$ DoF for its individual message and $0.2$ DoF for the common message, as desired. Note here that R1 can decode the common message ${\cal M}_0$ using just $U_{01}$, i.e., while decoding ${\cal M}_0$, $U_{02}$ can be regarded as noise; but, after decoding the common message, it can subtract contribution due to $U_{01}$ and $U_{02}$, since they are both generated by the two transmitters based on the same message. This is the key that allows us to realize the DoF benefit, even with independent transmit signaling.

Now, in the absence of common message, we can use the same scheme outlined above but with $U_{01} = U_{02} = 0$. With this change, we can achieve $0.6$ DoF for each individual message. Since at $\alpha = 0.6$, one can not attain more than $0.6$ DoF for the two individual messages (as per Theorem \ref{thm: per-user DoF wo common symmetric IC-CI journal}), our scheme with $U_{01} = U_{02} = 0$ is indeed GDoF-optimal, even in the absence of the common message. The signal level space of R1, in the absence of common message, is depicted in Fig. \ref{fig: exmple scheme no M0 ICCI journal}.

Let us focus on the decoding of message ${\cal M}_{1,pu}$ in the absence of common message. While decoding this message, R1 treats all other signals as noise. In this sense, ${\cal M}_{1,pu}$ experiences the signal-to-interference-plus-noise ratio (SINR) of $P^{0.4}$ (because the next interfering signal, namely, $U_{2,pu}$ is received at the SNR of $P^{0.6}$). However, only $0.2$ DoF are to be achieved for this message. This implies that a part of the receive signal space level space of R1 remains unused in the absence of the common message. Moreover, on comparing Figs. \ref{fig: exmple scheme with M0 ICCI journal} and \ref{fig: exmple scheme no M0 ICCI journal}, we observe that the common message produces a DoF improvement precisely by exploiting this unused portion.

In fact, it is clear that the unused DoF without the common message can be completely exploited by the common message to produce a DoF enhancement.

\section{Proof of Theorem \ref{thm: outer-bound IC-CI journal}} \label{app: proof of thm: outer-bound IC-CI journal}

Recall that the capacity region of the deterministic DM IC-CI, which can be considered as a generalization of the El Gamal-Costa deterministic IC \cite{Gamal_Costa}, is derived in \cite{jiang_xin_garg_IC_common_info}. The outer-bounds in this proof are inspired by the converse argument of \cite[Theorem 4]{jiang_xin_garg_IC_common_info} in the same manner that the outer bounds for the Gaussian IC in \cite{Etkin_Tse_Wang} are inspired from the converse arguments of Gamal and Costa \cite[Theorem 1]{Gamal_Costa}. However, the techniques in \cite[Proof of Theorem 4]{jiang_xin_garg_IC_common_info} must be suitably modified to account for the presence of noise. For instance, the outer-bounds here make use of different genie-aided side-information models than those considered in \cite[Proof of Theorem 4]{jiang_xin_garg_IC_common_info}. Furthermore, the analysis here is different from that of \cite{Etkin_Tse_Wang} because the transmit signals of the IC-CI can be arbitrarily correlated, unlike the case of the IC (without common message). Clearly, to derive a tight outer bound, it is necessary to limit, in some sense, the correlation between the transmit messages. This is done by noting that these signals are independent when conditioned on the common message (see Fact 3). In what follows, we detail how these general ideas are made to yield a tight outer-bound.

Let us first define $ V_1(t) \define h_{21} X_1(t) + Z_2(t) $ and $ V_2(t) \define h_{12} X_2(t) + Z_1(t) $
so that
\[
Y_1(t) = h_{11} X_1(t) + V_2(t) \quad \mbox{ and } \quad Y_2(t) = h_{22} X_2(t) + V_1(t).
\]
The following simple facts are used repeatedly in this proof.
\begin{fact}
Given ${\cal M}_0$ and ${\cal M}_i$, $X_i^n$ is deterministic for $i = 1,2$. Moreover, conditioned on ${\cal M}_0$, the pairs $({\cal M}_1,X_1^n)$ and $({\cal M}_2,X_2^n)$ are independent.
\end{fact}

\begin{fact}
Given ${\cal M}_0$, $V_i^n$ is independent of ${\cal M}_j$, $X_j^n$, and $V_j^n$ for $j \not= i$.
\end{fact}

\begin{fact}
Given ${\cal M}_0$, the transmit signals are independent.
\end{fact}

\begin{fact}
The additive noise $Z_i^n$ is independent of the messages and the transmit signals.
\end{fact}
Lastly, $\epsilon_n$ is a sequence such that $\epsilon_n \to 0$ as $n \to \infty$.

We next prove each bound separately starting with the derivation of the bound on $R_0+R_1$.

\underline{Proof of $R_0 + R_1 \leq \overline{G}_1'$ : } Recall, $\overline{G}_1'$ is given in Definition \ref{def: numbers Couter ICCI journal}.

By Fano's inequality, we have
\begin{eqnarray*}
\lefteqn{ n (R_0 + R_1 - \epsilon_n) \leq  I\Big( \mathcal{M}_0,\mathcal{M}_1;Y_1^n\Big) \leq  I \Big( \mathcal{M}_0,\mathcal{M}_1,\mathcal{M}_2;Y_1^n \Big) } \\
&& {} =  h \Big( Y_1^n \Big) - h \Big( Y_1^n \Big| \mathcal{M}_0,\mathcal{M}_1,\mathcal{M}_2 \Big) \\
&& {} =  h \Big( Y_1^n \Big) - h \Big( Y_1^n \Big| X_1^n, X_2^n,  \mathcal{M}_0,\mathcal{M}_1,\mathcal{M}_2 \Big) \quad (\mbox{using Fact 1}) \\
&& {} =  h \Big( Y_1^n \Big) - h \Big( Z_1^n \Big| X_1^n, X_2^n,  \mathcal{M}_0,\mathcal{M}_1,\mathcal{M}_2 \Big) \quad (\mbox{since differential entropy is translation invariant})\\
&& {} = h \Big( Y_1^n \Big) - h \Big( Z_1^n \Big) \quad (\mbox{using Fact 4}) \\
&& {} \leq  \sum_{t=1}^n \left\{ h \Big( Y_1(t) \Big) - h \Big( Z_1(t) \Big) \right\},
\end{eqnarray*}
where the last inequality follows since conditioning reduces differential entropy. Since translation does not change differential entropy, we may assume that $Y_1(t)$ is zero mean, $\forall$ $t$, which implies the same about $X_1(t)$ and $X_2(t)$. We will now prove that the variance of $Y_1(t)$ (denoted as ${\rm Var}(Y_1(t))$) is bounded, which allows us to bound its differential entropy. Consider the following:
\begin{eqnarray}
{\rm Var} \Big( Y_1(t) \Big) & = & \mathbb{E} \Big| Y_1(t) Y_1^*(t) \Big| \nonumber  \\
& = & 1 + |h_{11}|^2 \mathbb{E} |X_1(t)|^2 + |h_{12}|^2 \mathbb{E}|X_2(t)|^2 + 2 \cdot {\rm Re}\Big(h_{11}h_{12}^* \mathbb{E} |X_1(t) X_2^*(t)| \Big) \nonumber \\
& \leq & 1 + |h_{11}|^2 P_{1,t} + |h_{12}|^2 P_{2,t} + 2 |h_{11}| |h_{12}| \sqrt{P_{1,t} P_{2,t}}, ~\cdots~ \mathbb{E} |X_i(t)|^2 = P_{i,t}, \label{eq: bound ineq1 varY1 ICCI journal} \\
& \leq & 1 + |h_{11}|^2 P_{1,t} + |h_{12}|^2 P_{2,t} + 2 |h_{11}| |h_{12}| \frac{1}{2} \Big[ P_{1,t} + P_{2,t} \Big], \label{eq: bound ineq2 varY1 ICCI journal}
\end{eqnarray}
where ${\rm Re}(z)$ denotes the real part of the complex number $z$; inequality in (\ref{eq: bound ineq1 varY1 ICCI journal}) holds because (a) ${\rm Re}(z) \leq |z|$ for any $z \in \mathbb{C}$ and (b) the Cauchy-Schwartz inequality; and the one in (\ref{eq: bound ineq2 varY1 ICCI journal}) is true because the arithmetic mean is greater than or equal to the geometric mean. Since for a given variance, Gaussian distribution maximizes differential entropy, we get
\begin{eqnarray}
\lefteqn{ \frac{1}{n} \sum_{t=1}^n \left\{ h \Big( Y_1(t) \Big) - h \Big( Z_1(t) \Big) \right\} } \nonumber \\
&& {} \leq \frac{1}{n} \sum_{t=1}^n \log_2 \left\{ 1 + |h_{11}|^2 P_{1,t} + |h_{12}|^2 P_{2,t} + 2 |h_{11}| |h_{12}| \frac{1}{2} \Big[ P_{1,t} + P_{2,t} \Big] \right\} \nonumber \\
&& {} \leq \log_2 \left\{ 1 + |h_{11}|^2 \frac{1}{n} \sum_{t=1}^n P_{1,t} + |h_{12}|^2 \frac{1}{n} \sum_{t=1}^n P_{2,t} + 2 |h_{11}| |h_{12}| \frac{1}{n} \sum_{t=1}^n \frac{1}{2} \Big[ P_{1,t} + P_{2,t} \Big] \right\} \label{eq: bound G1prime ineq1 ICCI journal} \\
&& {} \leq \log_2 \left\{ 1 + |h_{11}|^2  + |h_{12}|^2  + 2 |h_{11}| |h_{12}| \right\} \label{eq: bound G1prime ineq2 ICCI journal} \\
&& {} = \log_2 \left\{ 1+ \big[ |h_{11}| + |h_{12}| \big]^2 \right\} = \overline{G}_1', \label{eq: bound G1prime ICCI journal}
\end{eqnarray}
where the inequality (\ref{eq: bound G1prime ineq1 ICCI journal}) holds due to Jensen's inequality; and inequality (\ref{eq: bound G1prime ineq2 ICCI journal}) is true because of the power constraint. Now, since $\epsilon_n \to 0$ as $n \to \infty$, we get
\[
R_0 + R_1 \leq \overline{G}_1',
\]
as desired.

\underline{Proof of $R_0 + R_2 \leq \overline{G}_2'$ : } follows by symmetry.

\underline{Proofs of $R_1 \leq \overline{D}_1$ and $R_2 \leq \overline{D}_2$ : } These bounds follow from the capacity of the point-to-point Gaussian channel.

\underline{Proof of $R_1 + R_2 \leq \overline{E}_1 + \overline{E}_2$ : } recall $\overline{E}_i$'s are given in Definition \ref{def: numbers Couter ICCI journal}. Note here that this bound looks identical to the one derived by Etkin et al for the IC \cite[Theorem 1]{Etkin_Tse_Wang}. We argue below that the proof of \cite[Theorem 1]{Etkin_Tse_Wang} is applicable to the IC-CI if the receivers are assumed to know the common message. Applying Fano's inequality, we obtain
\[
(R_1 + R_2 -\epsilon_n) \leq I \Big( {\cal M}_1;Y_1^n \Big| {\cal M}_0 \Big) + I \Big( {\cal M}_2;Y_2^n \Big| {\cal M}_0 \Big).
\]
Now, conditioned on ${\cal M}_0$, the transmit signals $X_1(t)$ and $X_2(t)$ are independent, and therefore, in the analysis henceforth, the IC-CI can just be regarded as the IC. Therefore, this bound can now be derived as in \cite[Theorem 1]{Etkin_Tse_Wang}. Consequently, we have (cf. \cite[equation (13)]{Etkin_Tse_Wang})
\[
n(R_1 + R_2 -\epsilon_n) \leq \sum_{t=1}^n \left\{ h\Big(Y_1(t) \Big| {\cal M}_0,V_1(t)\Big) - h \Big( Z_1(t) \Big) + h \Big( Y_2(t) \Big| {\cal M}_0,V_2(t) \Big) -h\Big(Z_2(t)\Big) \right\}
\]
and (cf. \cite[equation (14)]{Etkin_Tse_Wang})
\begin{equation}
\frac{1}{n}  \sum_{t=1}^n \left\{ h\Big(Y_i(t) \Big| {\cal M}_0,V_i(t)\Big) -h\Big(Z_i(t)\Big) \right\} \leq \overline{E}_i \label{eq: bound Ei ICCI journal}
\end{equation}
to obtain the desired result.

\underline{Proof of $R_0 + R_1 + R_2 \leq \overline{A}_2 + \overline{G}_1'$ : }

We apply Fano's inequality assuming that R2 knows $V_2^n$, $X_1^n$, and ${\cal M}_0$ to derive the following:
\begin{eqnarray}
\lefteqn{ n(R_0 + R_1 + R_2 - \epsilon_n) \leq I \Big( {\cal M}_1,{\cal M}_0;Y_1^n \Big) + I \Big( {\cal M}_2 ; Y_2^n,V_2^n,X_1^n,{\cal M}_0 \Big) } \nonumber \\
&& {} \hspace{-0.8cm} =  I \Big( {\cal M}_1,{\cal M}_0;Y_1^n \Big) + I \Big( {\cal M}_2 ; Y_2^n,V_2^n \Big| X_1^n,{\cal M}_0 \Big) \quad (\mbox{using Fact 1} ) \nonumber \\
&& {} \hspace{-0.8cm} =  h \Big( Y_1^n \Big) - h \Big( Y_1^n \Big| {\cal M}_1,{\cal M}_0 \Big) + h \Big( V_2^n \Big| {\cal M}_0, X_1^n \Big) - h \Big( V_2^n \Big| {\cal M}_0, {\cal M}_2, X_1^n \Big) \nonumber \\
&& {} \hspace{1cm} + h \Big( Y_2^n \Big| {\cal M}_0,X_1^n,V_2^n\Big) - h \Big( Y_2^n \Big| {\cal M}_0,{\cal M}_2, X_1^n, V_2^n \Big) \label{eq: key step R_0+R_1+R_2 ICCI journal} \\
&& {} \hspace{-0.8cm} = h \Big( Y_1^n \Big) -  h \Big( V_2^n \Big| {\cal M}_0 \Big) + h \Big( V_2^n \Big| {\cal M}_0 \Big) - h \Big( Z_1^n \Big) + h \Big( h_{22} X_2^n + Z_2^n \Big| {\cal M}_0, V_2^n \Big) - h \Big( Z_2^n \Big) \label{eq: key step 1 R_0+R_1+R_2 ICCI journal} \\
&& {} \hspace{-0.8cm} \leq \sum_{t=1}^n \left\{ h\Big( Y_1(t) \Big) - h\Big( Z_1(t) \Big) + h \Big( h_{22} X_2(t) + Z_2(t) \Big| {\cal M}_0, h_{12} X_2(t) + Z_1(t) \Big) - h \Big( Z_2(t) \Big)\right\}, \label{eq: key step 2 R_0+R_1+R_2 ICCI journal}
\end{eqnarray}
where the equality (\ref{eq: key step 1 R_0+R_1+R_2 ICCI journal}) holds due to Facts 1-4, and the subsequent inequality follows since conditioning reduces entropy.
From equation (\ref{eq: bound G1prime ICCI journal}), we have
\[
\frac{1}{n} \sum_{t=1}^n \left\{ h\Big( Y_1(t) \Big) - h\Big( Z_1(t) \Big) \right\} \leq \overline{G}_1'.
\]
It remains to bound $h \Big( h_{22} X_2(t) + Z_2(t) \Big| {\cal M}_0, h_{12} X_2(t) + Z_1(t) \Big)$. If $Y_2'(t) \define h_{22} X_2(t) + Z_2(t) $, then the covariance matrix of $Y_2'(t)$ and $V_2(t)$ is given by
\[
{\rm cov} \begin{bmatrix} Y_2'(t) \\ V_2(t) \end{bmatrix} = \begin{bmatrix} |h_{22}|^2 P_{2,t} + 1 & h_{22} h_{12}^* P_{2,t} \\ h_{22}^* h_{12} P_{2,t} &  |h_{12}|^2P_{2,t} + 1\end{bmatrix}, \\
\]
where $\mathbb{E} |X_2(t)|^2 = P_{2,t}$. Since the Gaussian distribution maximizes conditional differential entropy for a given covariance matrix, we obtain
\begin{eqnarray}
\lefteqn{ \frac{1}{n} \sum_{t=1}^n h \Big( h_{22} X_2(t) + Z_2(t) \Big| {\cal M}_0, h_{12} X_2(t) + Z_1(t) \Big) - h \Big( Z_2(t) \Big) } \nonumber \\
&& {} \leq \frac{1}{n} \sum_{t=1}^n \log \Big( 1 + \frac{|h_{22}|^2 P_{2,t}}{1+|h_{12}|^2 P_{2,t}} \Big) \leq \log \Big( 1 +  \frac{|h_{22}|^2 \frac{1}{n} \sum_{t=1}^n P_{2,t}}{1+ |h_{12}|^2 \frac{1}{n} \sum_{t=1}^n P_{2,t}} \Big) \nonumber \\
&& {} \leq \log \Big( 1 +  \frac{|h_{22}|^2}{1+ |h_{12}|^2 } \Big) = \overline{A}_2, \label{eq: bound A2 ICCI journal}
\end{eqnarray}
where the second inequality holds due to Jensen's inequality. Hence, we have $R_0 + R_1 + R_2 \leq \overline{A}_2 + \overline{G}_1'$.

\underline{Proof of $R_1 + R_2 \leq \overline{A}_2 + \overline{G}_1$ : } The proof of this bound is similar to that of the earlier one.

We apply Fano's inequality assuming that R1 knows ${\cal M}_0$ and R2 knows $V_2^n$, $X_1^n$, and ${\cal M}_0$ to derive the following:
\begin{eqnarray}
n(R_1 + R_2 - \epsilon_n) & \leq & I \Big( {\cal M}_1;Y_1^n,{\cal M}_0 \Big) + I \Big( {\cal M}_2 ; Y_2^n,V_2^n,X_1^n,{\cal M}_0 \Big)  \nonumber \\
&=&  h \Big( Y_1^n \Big| {\cal M}_0 \Big) - h \Big( Y_1^n \Big| {\cal M}_1,{\cal M}_0 \Big) + h \Big( V_2^n \Big| {\cal M}_0 \Big) - h \Big( V_2^n \Big| {\cal M}_0,{\cal M}_2 \Big). \nonumber \\
& \leq & \sum_{t=1}^n \left\{ h\Big( Y_1(t) \Big| {\cal M}_0 \Big) - h\Big( Z_1(t) \Big) + h \Big( h_{22} X_2(t) + Z_2(t) \Big| {\cal M}_0, h_{12} X_2(t) + Z_1(t) \Big) \right. \nonumber \\ && \hspace{6cm} - \left. h \Big( Z_2(t) \Big)\right\}. \nonumber
\end{eqnarray}
where the last inequality is obtained from the analysis that leads from (\ref{eq: key step R_0+R_1+R_2 ICCI journal}) to (\ref{eq: key step 2 R_0+R_1+R_2 ICCI journal}).
With inequality (\ref{eq: bound A2 ICCI journal}) already derived, it is sufficient to show that
$
\sum_{t=1}^n \left\{ h\Big( Y_1(t) \Big| {\cal M}_0 \Big) - h\Big( Z_1(t) \Big)  \right\} \leq \overline{G}_1. \label{eq: bound G1 ICCI journal}
$
The derivation of this bound is similar to that of (\ref{eq: bound G1prime ICCI journal}). The goal is to bound the variance of $Y_1(t)$, conditioned ${\cal M}_0$. Toward this end, we observe that the derivation of inequality (\ref{eq: bound ineq2 varY1 ICCI journal}) allows us to write
$
{\rm Var}\Big( Y_1(t) \Big| {\cal M}_0 \Big) \leq 1 + |h_{11}|^2 P_{1,t} + |h_{12}|^2 P_{2,t}
$
on noting that the transmit signals are independent conditioned on ${\cal M}_0$. Now, the analysis leading to bound (\ref{eq: bound G1 ICCI journal}) and the above bound on the conditional variance of $Y_1(t)$ together imply the desired inequality (\ref{eq: bound G1 ICCI journal}).

\underline{Proofs of Bounds $R_0 + R_1 + R_2 \leq \overline{A}_1 + \overline{G}_2'$ and $R_1 + R_2 \leq \overline{A}_1 + \overline{G}_2$ : } These follow by symmetry.

\underline{Proof of $R_0 + 2 R_1 + R_2 \leq \overline{A}_1 + \overline{G}_1' + \overline{E}_2$ : } 
Consider the following arguments:
\begin{eqnarray*}
\lefteqn{ n(R_0 + 2R_1 + R_2 - \epsilon_n) =  n\Big( [R_0+R_1] + R_1 + R_2 - \epsilon_n \Big)  } \nonumber \\
&& {} \leq I \Big( {\cal M}_0,{\cal M}_1;Y_1^n \Big) + I \Big( {\cal M}_1;Y_1^n,V_1^n,X_2^n,{\cal M}_0\Big) + I\Big({\cal M}_2;Y_2^n,{\cal M}_0,V_2^n\Big) \nonumber \\
&& {} = I \Big( {\cal M}_0,{\cal M}_1;Y_1^n \Big) + I \Big( {\cal M}_1;Y_1^n,V_1^n \Big| X_2^n, {\cal M}_0 \Big) +
I\Big({\cal M}_2;Y_2^n,V_2^n \Big| {\cal M}_0 \Big) \quad (\mbox{using Fact 1}) \nonumber \\
&& {} = h \Big( Y_1^n \Big) - h\Big( Y_1^n \Big|  {\cal M}_0,{\cal M}_1 \Big) + h \Big( V_1^n \Big| X_2^n, {\cal M}_0 \Big) - h \Big( V_1^n \Big| X_2^n, {\cal M}_0,{\cal M}_1 \Big) \nonumber \\
&& {} \hspace{2cm} + h \Big( Y_1^n \Big| V_1^n, X_2^n, {\cal M}_0 \Big) - h \Big( Y_1^n \Big| V_1^n, X_2^n, {\cal M}_0,{\cal M}_1 \Big) + h \Big( V_2^n \Big| {\cal M}_0 \Big) \nonumber \\
&& {} \hspace{2cm}  - h \Big( V_2^n \Big| {\cal M}_0, {\cal M}_2 \Big)  + h \Big( Y_2^n \Big| V_2^n, {\cal M}_0 \Big) - h \Big( Y_2^n \Big| V_2^n, {\cal M}_0, {\cal M}_2 \Big) \nonumber \\
&& {} = h \Big( Y_1^n \Big) -  h\Big( V_2^n \Big|  {\cal M}_0 \Big) +  h \Big( V_1^n \Big| {\cal M}_0 \Big) - h \Big( Z_2^n \Big) + h \Big( h_{11} X_1^n + Z_1^n \Big| V_1^n, {\cal M}_0 \Big) - h \Big( Z_1^n \Big) \nonumber \\
&& {} \hspace{2cm} + h \Big( V_2^n \Big| {\cal M}_0 \Big) - h \Big( Z_1^n \Big)  + h \Big( Y_2^n \Big| V_2^n, {\cal M}_0 \Big) - h \Big( V_1^n \Big| {\cal M}_0 \Big) \quad (\mbox{using Facts 1-4}) \nonumber \\
&& {} = h \Big( Y_1^n \Big) - h \Big( Z_1^n \Big)  +  h \Big( h_{11} X_1^n + Z_1^n \Big| V_1^n, {\cal M}_0 \Big) - h \Big( Z_1^n \Big)  +  h \Big( Y_2^n \Big| V_2^n, {\cal M}_0 \Big) -  h \Big( Z_2^n \Big) \\
&&{} \leq \sum_{t=1}^n \left\{ h \Big( Y_1(t) \Big) - h \Big( Z_1(t) \Big) \right\}  + \sum_{t=1}^n \left\{  h \Big( h_{11} X_1(t) + Z_1(t) \Big| V_1(t), {\cal M}_0 \Big) - h \Big( Z_1(t) \Big) \right\} \\
&& {} \hspace{2cm} + \sum_{t=1}^n \left\{ h \Big( Y_2(t) \Big| V_2(t), {\cal M}_0 \Big) -  h \Big( Z_2(t) \Big) \right\} \quad \mbox{ since conditioning reduces entropy}.
\end{eqnarray*}
Now the desired bound can be obtained by applying inequalities in (\ref{eq: bound G1prime ICCI journal}), (\ref{eq: bound A2 ICCI journal}), and (\ref{eq: bound G1 ICCI journal}).

\underline{Proof of $2 R_1 + R_2 \leq \overline{A}_1 + \overline{G}_1 + \overline{E}_2$ : }

The analysis is almost similar to that of the bound $R_0 + 2 R_1 + R_2 \leq \overline{A}_1 + \overline{G}_1' + \overline{E}_2$. Recall that we modified the proof of the bound $R_0+R_1+R_2 \leq \overline{A}_2 + \overline{G}_1'$ to derive the bound $R_1+R_2 \leq \overline{A}_2 + \overline{G}_1$. In an analogous fashion, we can modify the proof of bound $R_0 + 2 R_1 + R_2 \leq \overline{A}_1 + \overline{G}_1' + \overline{E}_2$ to obtain this bound.

\underline{Proofs of Bounds $R_0 + R_1 + 2 R_2 \leq \overline{A}_2 + \overline{G}_2' + \overline{E}_1$ and $R_1 + 2R_2 \leq \overline{A}_2 + \overline{G}_2 + \overline{E}_1$ : } These follow by symmetry.

\section{Proofs of Theorems \ref{thm: cap within 1 bit ICCI journal} and \ref{thm: GDoF region ICCI journal}} \label{app: proof of thm: cap within 1 bit ICCI journal thm: GDoF region ICCI journal}
The two proofs are provided in the following two sub-sections.

\subsection{Proof of Theorem \ref{thm: cap within 1 bit ICCI journal}} \label{app: proof of thm: cap within 1 bit ICCI journal}

Since the inner and outer bounds have the same shape, it is possible to perform a bound-by-bound analysis to prove that the gap between the two is at most one bit. Suppose $\Delta_{R_i}$ be the difference between the outer bound in $R_i$, which is equal to $\overline{D}_i$ and inner bounds on it, which is $D_i$, i.e., set $\Delta_{R_i} = \overline{D}_i - D_i$ (see Definitions of ${\bf C}_{\rm inner}(H)$ and ${\bf C}_{\rm outer}(H)$). Similarly, we define $\Delta_{R_0+R_i}$, $\Delta_{R_1+R_2}$, $\Delta_{R_0+R_1+R_2}$, $\Delta_{2R_i+R_j}$, and $\Delta_{R_0+2R_i+R_j}$ as follows: For each $i \in \{1,2\}$, if $j \in \{1,2\}$ such that $i \not= j$, then
\begin{eqnarray*}
\Delta_{R_i} & \define & \overline{D}_i - D_i,   \\
\Delta_{R_0+R_i} & \define & \overline{G}_i'-G_i, \\
\Delta_{R_1+R_2} & \define & \min \left\{ \overline{E}_1+\overline{E}_2 , ~ \overline{A}_1+\overline{G}_2, ~ \overline{A}_2 + \overline{G}_1 \right\} - \min \left\{  E_1+E_2, ~ A_1 + G_2, ~ A_2 + G_1 \right\} \\
& \leq & \max\left\{ \overline{E}_1+\overline{E}_2 - E_1-E_2, ~ \overline{A}_1+\overline{G}_2 - A_1 - G_2, ~ \overline{A}_2 + \overline{G}_1 - A_2 - G_1 \right\}, \\
\Delta_{R_0+R_1+R_2} & \define & \min \left\{  \overline{A}_1+\overline{G}_2', ~ \overline{A}_2 + \overline{G}_1' \right\} - \min \left\{ A_1 + G_2', ~ A_2 + G_1' \right\} \\
& \leq & \max \left\{ \overline{A}_1+\overline{G}_2' - A_1 - G_2', ~ \overline{A}_2 + \overline{G}_1' - A_2 - G_1' \right\}, \\
\Delta_{2R_i+R_j} & \define & \overline{A}_i + \overline{G}_i + \overline{E}_j - A_i - G_i - E_j, \\
\Delta_{R_0+2R_i+R_j} & \define & \overline{A}_i + \overline{G}_i' + \overline{E}_j - A_i - G_i' - E_j.
\end{eqnarray*}

Suppose the following inequalities for each $i \in \{1,2\}$ are true:
\begin{eqnarray}
\Delta A_i & \define & \overline{A}_i - A_i < 1, \quad \Delta D_i \define \overline{D}_i - D_i < 1, \quad \Delta E_i  \define  \overline{E}_i - E_i < 1, \label{eq: delta Ai ICCI journal}  \\
\Delta G_i & \define & \overline{G}_i - G_i \leq 1, \quad \Delta G_i' \define \overline{G}_i' - G_i' < 2. \label{eq: delta Gi' ICCI journal}
\end{eqnarray}
Then, it can be easily verified that for each $i \in \{1,2\}$, if $j \in \{1,2\}$ such that $i \not= j$, then
\begin{eqnarray*}
\Delta_{R_i} < 1, ~ \Delta_{R_0+R_i} < 2, ~ \Delta_{R_1+R_2} < 2, ~ \Delta_{R_0+R_1+R_2} < 3, ~ \Delta_{2R_i+R_j} < 3, ~ \Delta_{R_0+2R_i+R_j} < 4,
\end{eqnarray*}
which together imply that the per-coordinate gap between the inner and outer bounds is at most one bit, for all values of $H$.

Thus, it remains to prove the five inequalities in (\ref{eq: delta Ai ICCI journal})-(\ref{eq: delta Gi' ICCI journal}). All these inequalities can be proved in a similar manner. We prove below the last one with $i=1$. We first lower-bound $G_1'$. By definition of $x_{12}$, $|h_{12}|^2 x_{12} \leq 1$. Therefore,
\[
G_1' = \log_2 \frac{1+ |h_{11}|^2 + |h_{12}|^2}{1 +|h_{12}|^2 x_{12}} \geq \log_2 \left( 1+|h_{11}|^2 + |h_{12}|^2 \right) - 1.
\]
Hence,
\begin{eqnarray*}
\Delta G_1' & \leq & \log_2  \frac{1 + |h_{11}|^2 + |h_{12}|^2 + 2|h_{11}| |h_{12}|}{1+|h_{11}|^2+|h_{12}|^2} +1 \\
& = & \log_2 \left( 1 + \frac{2 |h_{11}| |h_{12}|}{1+|h_{11}|^2+|h_{12}|^2} \right) +1 \\
& < & 2,
\end{eqnarray*}
where the last inequality holds since $ 1  >  \frac{2 |h_{11}| |h_{12}|}{1+|h_{11}|^2+|h_{12}|^2} $ (because $ 1 + (|h_{11}| - |h_{12}|)^2  >0$).

\subsection{Proof of Theorem \ref{thm: GDoF region ICCI journal}} \label{app: proof of thm: GDoF region ICCI journal}

Since the inner and outer-bounds are within one bit for all values of $H$, we can compute the GDoF region by treating the outer-bound ${\bf C}_{\rm outer}(H)$ as the capacity region. That is, we have
\begin{eqnarray*}
\lefteqn{ \mathbf{D}(\overline{\mathbf{\alpha}}) = \Biggl\{ (d_0,d_1,d_2) \in \mathbb{R}_+^3 \biggl| ~\mbox{for a } P > 0, ~  |h_{ij}|^2 = P^{\alpha_{ij}} ~\! \forall ~i,j\in\{1,2\} \biggr. \Biggr.  } \\
&& {} \hspace{2cm} \Biggl. \mbox{ and } (R_0,R_1,R_2) \in \mathbf{C}_{\rm outer}(H) \mbox{ such that } d_k = {\rm MG}(R_k) ~\! \forall k = 1,2,3 \Biggr\},
\end{eqnarray*}
where ${\rm MG}(x) = \lim_{P \to \infty} \frac{x}{\log_2 P}$. Note we have ${\bf C}_{\rm outer}(H)$ in place of the capacity region ${\bf C}(H)$. Thus, to prove this theorem, it is sufficient to prove that
\begin{equation}
{\rm MG}(\overline{A}_1) = {\sf a_1}, ~ {\rm MG}(\overline{D}_1) = {\sf d_1}, ~ {\rm MG}(\overline{E}_1) = {\sf e_1}, ~ \mbox{and } {\rm MG}(\overline{G}_1) = {\rm MG}(\overline{G}_1') = {\sf g_1}. \label{eq: DoF calculation ICCI journal}
\end{equation}
While the above equalities can be easily verified, we provide below the details of one of them, namely, $ {\rm MG}(\overline{G}_1') = {\sf g_1}$:
\begin{eqnarray*}
\overline{G}_1' & = & \mathbf{\sf C}\left( [|h_{11}+ h_{12}|^2]\right) = \log_2 \left(1 + |h_{11}|^2 + |h_{12}|^2 + 2 |h_{11}| \cdot |h_{12}| \right) \\
& = & \log_2 \left( 1 + P^{\alpha_{11}} + P^{\alpha_{12}} + P^{\frac{1}{2}(\alpha_{11}+\alpha_{12})} \right) \\
\Longrightarrow \; {\rm MG}(\overline{G}_1') = {\sf g_1} & = & \max \left\{ \alpha_{11}, \alpha_{12}, \frac{1}{2} \left( \alpha_{11}+\alpha_{12}\right) \right\} = \max \left\{ \alpha_{11}, \alpha_{12} \right\}.
\end{eqnarray*}
Similarly, we can prove all equalities in (\ref{eq: DoF calculation ICCI journal}) to complete the proof of this theorem.

\section{Proof of Theorem \ref{thm: per-user DoF symmetric IC-CI journal}} \label{app: proof of thm: per-user DoF symmetric IC-CI journal}

For the symmetric Gaussian IC-CI, the GDoF region simplifies to
\begin{eqnarray*}
\lefteqn{ \mathbf{D}_{sym}(\alpha) = \Big\{ (d_0,d_1,d_1) \Big| ~ d_0, d_1\geq 0; \Big. \Big. } \\
&& {} \hspace{2cm} d_0 + d_1\leq \max \big\{1, \alpha \big\}; \\
&& {} \hspace{2cm} d_1 \leq \min \Big( 1, \max \big\{ \alpha,1-\alpha \big\}\Big) ; \\
&& {} \hspace{2cm} \Big. d_0 + 2d_1 \leq \max \big\{1,\alpha\big\} + (1-\alpha)^+ \Big\}.
\end{eqnarray*}
where we have made use of the equalities $d_1 = d_2$, $\alpha_{11} = \alpha_{22} = 1$, and $\alpha_{12} = \alpha_{21} = \alpha$; and for simplicity, we have denoted the GDoF region by $\mathbf{D}_{sym}(\alpha)$, instead of $\mathbf{D}(\overline{\alpha})$.

Define $ {\bf d}^{\star}(\alpha)$ to be equal to the function on the right hand of equation (\ref{eq: thm: per-user DoF symmetric IC-CI journal}). Hence, we need to prove that ${\bf d}_{\sf ICCI}(\alpha) = {\bf d}^{\star}(\alpha)$. Further, let ${\bf d}_{\sf IC}(\alpha)$ be as in equation (\ref{eq: thm: per-user DoF wo common symmetric IC-CI journal}) of Theorem \ref{thm: per-user DoF wo common symmetric IC-CI journal}; and define ${\bf d}^{\star}_0(\alpha) \define 2 \Big( {\bf d}^{\star}(\alpha) - {\bf d}_{\sf IC}(\alpha) \Big)$ so that ${\bf d}^{\star}_0(\alpha)$ is twice the function on the right hand side of equation (\ref{eq: increase thm: per-user DoF symmetric IC-CI journal}). Alternatively, ${\bf d}^{\star}(\alpha) = \frac{1}{2}{\bf d}^{\star}_0(\alpha) + {\bf d}_{\sf IC}(\alpha)$.

Note from the definitions of ${\bf d}_{\sf ICCI}(\alpha)$ and $\mathbf{D}_{sym}(\alpha)$ that
\begin{equation}
{\bf d}_{\sf ICCI}(\alpha) = \max_{(d_0,d_1,d_1) \in \mathbf{D}_{sym}(\alpha)} ~ \frac{1}{2} \left( d_0 + 2 d_1 \right). \label{eq: alt defn of dICCI journal}
\end{equation}

To show the equality ${\bf d}_{\sf ICCI}(\alpha) = {\bf d}^{\star}(\alpha)$, we prove, respectively, in the following two sub-sections that ${\bf d}_{\sf ICCI}(\alpha) \geq {\bf d}^{\star}(\alpha)$ and ${\bf d}_{\sf ICCI}(\alpha) \leq {\bf d}^{\star}(\alpha)$, which together imply the desired equality.

\subsection{Proof of Inequality ${\bf d}_{\sf ICCI}(\alpha) \geq {\bf d}^{\star}(\alpha)$}

With simple substitution, we observe that
\begin{equation}
\Big( {\bf d}^{\star}_0(\alpha), {\bf d}_{\sf IC}(\alpha), {\bf d}_{\sf IC}(\alpha) \Big) \in \mathbf{D}_{sym}(\alpha), \label{eq: inclusion of key point sym per-user w/ common ICCI journal}
\end{equation}
which implies ${\bf d}_{\sf ICCI}(\alpha) \geq \frac{1}{2}{\bf d}^{\star}_0(\alpha) + {\bf d}_{\sf IC}(\alpha) = {\bf d}^{\star}(\alpha)$. In fact, the point $\Big( {\bf d}^{\star}_0(\alpha), {\bf d}_{\sf IC}(\alpha), {\bf d}_{\sf IC}(\alpha) \Big)$ lies on the boundary of $\mathbf{D}_{sym}(\alpha)$, i.e.,
\begin{equation}
(d_0,d_1,d_1) \in \mathbf{D}_{sym}(\alpha) ~ \mbox{ and } ~ d_1 = {\bf d}_{\sf IC}(\alpha) ~ \Rightarrow ~ d_0 \leq {\bf d}^{\star}_0(\alpha). \label{eq: proof of dICCI key fact}
\end{equation}

\subsection{Proof of Inequality ${\bf d}_{\sf ICCI}(\alpha) \leq {\bf d}^{\star}(\alpha)$}

Suppose $\left( {d}_0^{opt}(\alpha),d_1^{opt}(\alpha),d_1^{opt}(\alpha) \right)$ be the argument of the maximization in (\ref{eq: alt defn of dICCI journal}), i.e.,
\begin{eqnarray*}
\Big( {d}_0^{opt}(\alpha),d_1^{opt}(\alpha),d_1^{opt}(\alpha) \Big) \define {\rm arg}  \max_{(d_0,d_1,d_1) \in \mathbf{D}_{sym}(\alpha)} ~ \frac{1}{2} \left( d_0 + 2 d_1 \right)
\end{eqnarray*}
and
\[
{\bf d}_{\sf ICCI}(\alpha) = \frac{1}{2} {d}_0^{opt}(\alpha) + d_1^{opt}(\alpha).
\]

If suppose $d_1^{opt}(\alpha) = {\bf d}_{\sf IC}(\alpha)$, then ${d}_0^{opt}(\alpha) = {\bf d}^{\star}_0(\alpha)$ because of the implication in (\ref{eq: proof of dICCI key fact}), which immediately implies that ${\bf d}_{\sf ICCI}(\alpha) = \frac{1}{2}{\bf d}^{\star}_0(\alpha) + {\bf d}_{\sf IC}(\alpha) = {\bf d}^{\star}(\alpha)$, and hence the theorem. Therefore, let us consider the remaining case, where $d_1^{opt}(\alpha) \not= {\bf d}_{\sf IC}(\alpha)$. Note, by definition of ${\bf d}_{\sf IC}(\alpha)$, that
\[
{\bf d}_{\sf IC}(\alpha) = \max_{(d_0,d_1,d_1) \in \mathbf{D}_{sym}(\alpha)} d_1.
\]
Hence, in the following, we may consider that $d_1^{opt}(\alpha) < {\bf d}_{\sf IC}(\alpha)$. Let $\epsilon \define {\bf d}_{\sf IC}(\alpha) - d_1^{opt}(\alpha)$. Now since ${\bf d}_{\sf ICCI}(\alpha) \geq {\bf d}^{\star}(\alpha)$, we have
\begin{eqnarray*}
\lefteqn{ \frac{1}{2} {d}_0^{opt}(\alpha) + d_1^{opt}(\alpha) \geq \frac{1}{2}{\bf d}^{\star}_0(\alpha) + {\bf d}_{\sf IC}(\alpha) } \\
&& {} \Rightarrow {d}_0^{opt}(\alpha) - 2 \epsilon \geq {\bf d}^{\star}_0(\alpha) + 2 \Big( {\bf d}_{\sf IC}(\alpha) - d_1^{opt}(\alpha) \Big) - 2 \epsilon \\
&& {} \Rightarrow {d}_0^{opt}(\alpha) - 2 \epsilon \geq {\bf d}^{\star}_0(\alpha) \geq 0.
\end{eqnarray*}
Hence, from the definition of $\mathbf{D}_{sym}(\alpha)$, we observe that
\begin{eqnarray*}
\lefteqn{ \Big( {d}_0^{opt}(\alpha),d_1^{opt}(\alpha),d_1^{opt}(\alpha) \Big)  \in \mathbf{D}_{sym}(\alpha) } \\
&& {} \Longrightarrow \; P(\alpha) \equiv \Big( {d}_0^{opt}(\alpha)-2\epsilon,d_1^{opt}(\alpha)+\epsilon,d_1^{opt}(\alpha)+\epsilon \Big) \in \mathbf{D}_{sym}(\alpha).
\end{eqnarray*}
Therefore,
\[
\left. \left(d_1 + \frac{d_0}{2}\right) \right|_{P(\alpha)} = d_1^{opt}(\alpha) + \frac{d_0^{opt}(\alpha)}{2} = {\bf d}_{\sf ICCI}(\alpha).
\]
Moreover, since $d_1$-coordinate of $P(\alpha)$ is equal to ${\bf d}_{\sf IC}(\alpha)$, (\ref{eq: proof of dICCI key fact}) implies that $d_0$-coordinate of $P(\alpha)$ can at most be equal to ${\bf d}^{\star}_0(\alpha)$. Hence,
\[
\left. \left(d_1 + \frac{d_0}{2}\right) \right|_{P(\alpha)} = {\bf d}_{\sf ICCI}(\alpha) \leq \frac{1}{2} {\bf d}^{\star}_0(\alpha) + {\bf d}_{\sf IC}(\alpha) = {\bf d}^{\star}(\alpha),
\]
as desired.

\section{Conclusion}      \label{sec: conclusion ICCI journal}

Explicit inner and outer bounds to the capacity region of the Gaussian IC-CI are determined and shown to be within a universal bounded gap of one bit, independently of channel parameters. Remarkably, the simple achievable scheme whose achievable rate region is the inner bound involves independent signaling at the transmitters which implies that it entirely forgoes the opportunity for transmitter cooperation that exists due to the shared knowledge of the common message at both transmitters. Nevertheless, through a characterization of the generalized degree of freedom region of the Gaussian IC-CI, it is shown that the presence of common message can lead to a very substantial (possibly unbounded) improvement in the total achievable rate over that achievable over the usual interference channel without a common message, to the extent that even the degrees of freedom achievable per user increase. An intuitive explanation of this DoF improvement is provided through which it is seen that sending just individual messages over the interference channel fundamentally doesn't fully exploit the available signal level dimensions at the receivers but the transmission of a common message allows for the full use of the potential of same physical (interference) channel.

\bibliographystyle{IEEEtran}
\bibliography{v5_journal_IC-CI.bbl}
\end{document}